\documentclass[superscriptaddress, onecolumn]{revtex4-2}

\usepackage{xurl}      
\usepackage[hypertexnames=false, breaklinks=true]{hyperref}
\usepackage{graphicx}
\usepackage{geometry}
\usepackage{multirow}
\usepackage{lipsum} 
\usepackage{tcolorbox}
\usepackage{bm}
\usepackage{xr}
\usepackage{float}
\usepackage{amsmath}
\usepackage{amssymb}
\usepackage{amsthm}
\usepackage{amsfonts}
\usepackage{booktabs}
\usepackage{xcolor}
\usepackage{soul}
\usepackage{svg}
\usepackage[normalem]{ulem}
\usepackage{booktabs}
\usepackage{xspace}
\usepackage{algorithm}
\usepackage{algpseudocode}
\usepackage{babel}
\usepackage[hypertexnames=false]{hyperref}

\begin{document}

\title{An analytical and experimental study of the energy transition discourse on YouTube}

\author{Aleix Bassolas}
\affiliation{Eurecat, Centre Tecnològic de Catalunya, Barcelona, Spain}
\author{Piero Birello}
\affiliation{Eurecat, Centre Tecnològic de Catalunya, Barcelona, Spain}
\affiliation{Department of Network and Data Science, Central European University, Vienna, Austria.}
\author{Julian Vicens}
\affiliation{Eurecat, Centre Tecnològic de Catalunya, Barcelona, Spain}

\begin{abstract}
Energy production and management face significant political, economic, and environmental challenges, yet the rise in information consumption through social media undermines the availability of reliable knowledge to the general public. This study examines the ideas discussed in the energy transition content on YouTube, assesses the most effective methods of communicating knowledge and information to the general public, and identifies the most engaged audiences. We examine videos related to the subject, analysing the themes discussed, the language used, and the emotions conveyed on YouTube, linking language formality to user engagement. To test the relationship experimentally, we uploaded original content to YouTube through two mirror channels containing the same material but using different levels of language formality. Our results indicate that conversational content reaches a broader audience, but retention rates are higher on the academic channel beyond the initial video segments. Interest in the topic varies by viewer profile, with younger individuals and women showing greater engagement regardless of language style.
\end{abstract}

\maketitle

\section*{Introduction}
Energy is a fundamental resource in modern society, powering transportation, industry, waste management, and telecommunications \cite{hasanuzzaman2020energy}. In recent years, global energy consumption has risen due to population growth and increasing demand from emerging economies \cite{van2019amplification}. However, the increase in energy production comes with negative externalities, including environmental degradation and the depletion of natural resources \cite{friedrich1993external, malla2009co2, sakulniyomporn2011external, rodrigues2020drivers}. Therefore, transitioning to a more sustainable energy system that reduces emissions, increases the share of renewable energy, and curtails overall energy consumption is crucial \cite{solomon2011coming}

The energy transition faces several challenges, including technological \cite{ostergaard2021recent}, political \cite{lee2019global}, and cultural \cite{carley2018adaptation}. While renewable energy sources have seen increasing adoption worldwide, their integration with existing production systems and technologies remains a significant challenge \cite{markard2018next}. On the political front, recent conflicts have disrupted energy resource availability and influenced policy shifts \cite{kuzemko2022russia}. Reducing overall energy consumption requires substantial changes in citizen behaviour \cite{steg2015understanding} and cultural attitudes \cite{komendantova2020discourses, komendantova2021transferring}, fostering a more responsible approach to energy use. However, raising public awareness about the importance of sustainability and encouraging energy-efficient behaviours remain major challenges.

Thus, an effective and reliable scientific communication becomes crucial to have an informed society and shape cultural norms. However, with the rise in information consumption fuelled by the adoption of social media, gaining visibility and actively engaging citizens has become increasingly difficult \cite{althaus2000patterns, fletcher2017impact}. Since the rise of online social networks, researchers have sought to understand which content features influence user interest and engagement. On Twitter (now X), studies have shown that factors such as the number of hashtags, URLs, and followers positively impact content reach and engagement \cite{jenders2013analyzing,nesi2018assessing}. Similarly, on YouTube, the number of subscribers and the upload date play a key role in predicting future views \cite{hoiles2017engagement}. Content type also affects the spread of information \cite{lagnier2013predicting,guille2013information,li2017survey}, videos in the music and entertainment categories tend to generate higher user engagement \cite{hoiles2017engagement}. Beyond content type, the emotions and sentiments expressed in posts influence both diffusion and engagement. Content with strong emotional appeal, whether positive or negative, tends to attract more interest \cite{dubovi2021interactions}. Some studies suggest that while negative sentiment narratives faster user reactions, positive sentiment promotes longer-term content diffusion \cite{ferrara2015quantifying}. Additionally, content with a negative sentiment has been associated with increased views and comment activity on YouTube \cite{munaro2021engage}. In fact, YouTube has become a widely used platform for educational content, offering longer videos than other platforms, which allows for the discussion of more complex subjects \cite{shoufan2022youtube}. While recent studies have examined engagement with science-related videos \cite{welbourne2016science, yang2022science}, the role of language formality in this context remains under-explored and has provided inconclusive results \cite{gretry2017don,munaro2024does}. 

This study has a two-fold objective: to understand how information on the energy transition is conveyed on YouTube and whether language features, specifically formality, influence engagement. To do so, we analyse the topics and ideas in energy transition content on YouTube, examine the language features in terms of emotions and formality, and conduct an experiment to investigate the influence of language in audience engagement and profile. In the first part of this study, we collected videos related to energy transition keywords and analysed the topics covered, the sentiments expressed, and their correlation with engagement metrics (views, likes, and comments). In the second part, we produced 20 original science communication videos and published them on two identical channels that differed only in language formality. We then conducted systematic comparisons and statistical tests to evaluate the effect of language formality on content dissemination, focusing on the engagement metrics (views, likes, and comments) and user retention. We also identified the audience segments that are more interested in the content.

\section*{Materials and Methods}

\subsection*{Data}

We conducted two complementary studies on energy transition content on YouTube: one focused on analysing existing videos, and the other on examining original content produced and published on the platform. Data for both studies were collected using the YouTube Data API \cite{googleYouTubeData}, following the official documentation and adhering to the platform’s quota limitations. To capture different aspects of public discourse on energy management and the energy transition, we selected 13 key concepts: energy economy, energy supply and demand, geopolitics of energy, electricity market, renewable energy, energy and mobility, energy saving, energy storage, energy transition, energy efficiency, decolonisation, energy model, and energy resources. For each concept, we retrieved information about 200 Spanish videos. The data was collected using the \textit{Search} function \cite{googleSearchYouTube} of the YouTube Data API, which allows content searches based on specific keywords. The search was performed using relevant keywords for each concept, specifying Spanish in the \textit{relevanceLanguage} field. However, while this parameter prioritises a given language, it does not guarantee that all retrieved content is in the intended language. We have detected the language of the video descriptions and transcriptions, keeping only those in Spanish. Additionally, to ensure that the videos were meaningful to the subject, we ensured that the video descriptions included variations of the terms energy or electricity. The final dataset analysed has 2,108 videos distributed across the 13 concepts. Since the algorithm governing video retrieval is not publicly disclosed, we cannot provide further details on how videos are selected. We provide a detailed analysis of the dataset in Appendix \ref{Appendix_S1}.

The data from the channels created during this research was collected using two sources: the YouTube API and the Google Ads API. The data from YouTube was obtained through the Analytics and Reporting API \cite{googleYouTubeAnalytics}, which provides access to daily and aggregated measures of engagement and retention to channel owners \cite{googleMetricsYouTube}. Engagement measures include the number of views, likes, and comments, and retention measures include the \textit{audience watch ratio} and the \textit{relative retention performance}. We collected data for the 20 challenges for which we published videos. The first 2 videos were used to assess the engagement on YouTube for content posted on newly created channels. The following 3 videos were used to test the Google Ads platform and adjust the campaign parameters (available in Appendix \ref{Appendix_S2}). Thus, the main analysis is centred on the last 15 challenges. Considering that each challenge has four different formats and there are two mirror channels, we analysed a total of 120 videos. Each challenge had two complete videos, which were longer, and two brief videos (See Fig. \ref{fig:duration_distribution}). The complete videos have a promoted and a non-promoted version, and the brief videos, both promoted, have standard and short formats with vertical orientation. The detailed data related to the Google Ads promotion campaigns is available through the Google Ads API \cite{googleOverviewGoogle}. Besides the number of visualisations and user interactions (clicks), the platform also provides the number of impressions. The metrics are also provided disaggregated by age, gender, and device. The data from Google Ads is at the campaign level, grouping the videos for each individual promotion. Thus, we have collected and analysed data on 15 challenges.

\subsection*{Methods}

\subsubsection*{Topic Modelling}

Topic modelling is a widely used NLP technique that identifies latent themes within a collection of documents, in our case, video transcripts. The topics are represented by a group of words that are more likely to be found together in the same document and likely belong to a shared semantic area. To facilitate the extraction of meaningful topics, we have preprocessed the texts using the package Spacy \cite{vasiliev2020natural} by removing stopwords, pronouns, and verbs, keeping names, proper names, adjectives, and adverbs.
We have used the model Latent Dirichlet Allocation (LDA) to detect the topics in our corpus of video transcripts, which is a probabilistic model based on Bayesian methods. The model extracts topics from a collection of documents, assuming that each document is a mixture of topics and each topic comes from a distribution of words that is learned through statistical inference \cite{latorre2021topic,jelodar2021nlp}. 

\subsubsection*{Polarity calculation}

VADER model is built on several preexisting and validated word classification dictionaries and assigns one or multiple categories to words. The assignment has real-valued scores that capture the intensity of sentiment conveyed by a word and, therefore, content. It is specifically designed for social networks where the textual content contains informal language, slang, abbreviations, or emoticons. VADER provides three polarity scores (positive, neutral, negative) proportionally so that all the scores add up to 1.

\subsubsection*{General Inquirer}

The second tool we used is called General Inquirer \cite{stone1963computer}, based on a dictionary that includes a list of word categories related to emotions, cognitive or psychological processes, among others. The details of the categories we used are available in the Supplementary Information.

\subsubsection*{Formality calculation}

 To measure the formality of a text, we employed the XLM-Roberta-base \cite{NEURIPS2019_c04c19c2} language model trained on the X-FORMAL \cite{li2022deep} dataset from the English-translated transcripts. The model receives as input text and returns two complementary scores, formal and informal. It is susceptible to punctuation and capitalisation since it is designed for written text, yet YouTube transcripts contain neither. To handle this, we detected sentences within the transcripts, introduced periods to separate them, and capitalised the first letter of each sentence. We calculated formality for each sentence separately and then averaged across all sentences in a transcript.

\section*{Results}
\subsection*{An analysis of the discussion on YouTube of the energy}

We first conducted a statistical analysis of energy-related content within the Spanish YouTube community (Fig. \ref{fig:statistics_distribution} and Fig. \ref{fig:statistics_concept}). The results indicate that total views have the highest engagement values, followed by likes and comments. Comparing the engagement across the 13 concepts, the content related to the \textit{renewable energy} has the highest number of views and comments. We also calculated the ratio between the average number of comments and views to assess the user interaction per view, where concepts with higher values are the \textit{electric market} and the \textit{energy model}.

\begin{figure}[!htbp]
\begin{center}
\includegraphics[width=1.1\textwidth]{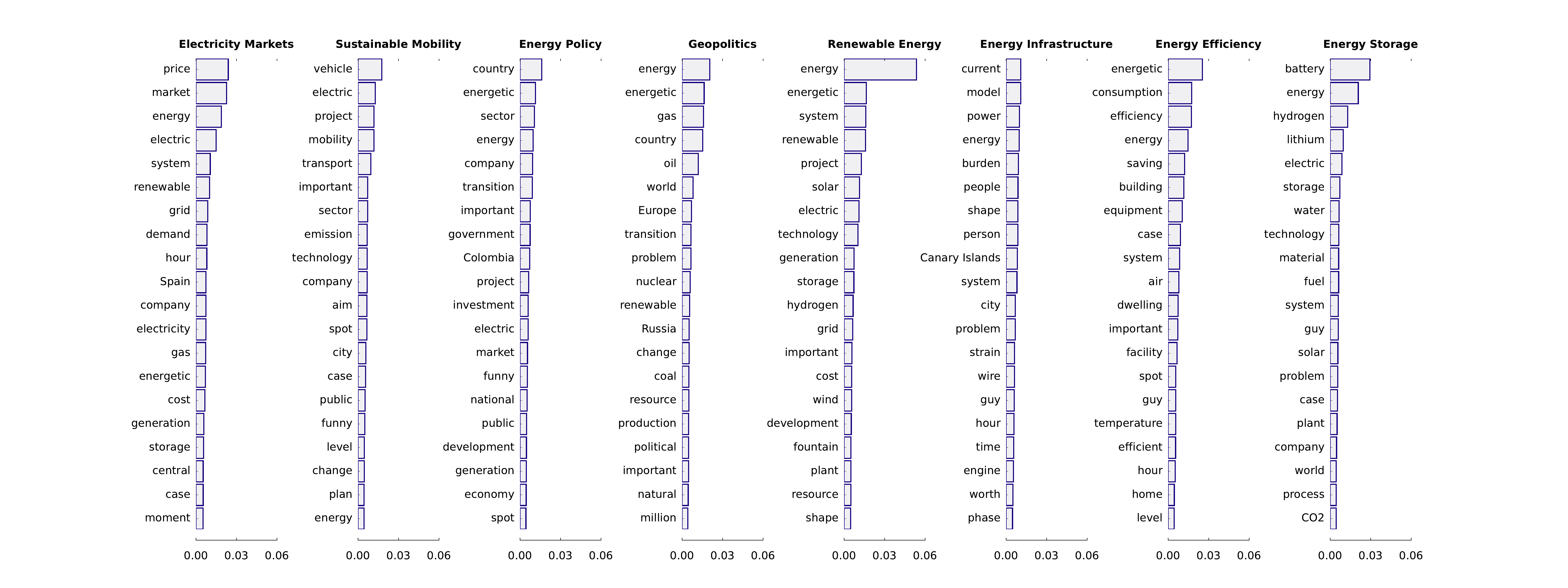}
\end{center}
\caption{\textbf{Importance of the words in the detected topics.} Ranking and weight of the most important words by topic detected.}
\label{fig:topic_words}
\end{figure}

We have conducted an LDA topic modelling to identify the main themes discussed in YouTube videos and how the concepts are distinguished. The eight topics detected are depicted in Fig. \ref{fig:topic_words}, where a weighted set of words represents each topic. Despite the concepts revolving around the same shared subject, the topics have distinctive words. 

Each video was assigned a single topic based on the highest probability score from the topic modelling process. The eight identified topics reflect distinct thematic areas within the Spanish YouTube energy discourse:

\begin{itemize}
    \item \textit{Electricity markets} includes content related to the electricity market, energy models, supply and demand, and the energy economy. It captures discussions around the economic and regulatory aspects of energy systems.
    \item \textit{Sustainable mobility} focuses on transportation-related terms such as vehicle and transport, highlighting the role of mobility in the energy transition.
    \item \textit{Energy policy} encompasses videos discussing legislation, regulation, and government strategies, addressing how political frameworks shape the energy sector.
    \item \textit{Geopolitics} features mentions of countries and energy resources like oil and gas, reflecting the international and strategic dimensions of energy.
    \item \textit{Renewable energy} contains the largest number of videos, likely due to its broader and more accessible terminology with words such as electric, energy, solar, and wind. This topic covers interconnected themes like renewable energy, energy storage, and energy resources.
    \item \textit{Energy infrastructure} relates to physical systems and projects supporting energy generation and distribution, including terms like grid, installation, and development.
    \item \textit{Energy efficiency} groups content around terms such as energy savings, buildings, decarbonization, and efficiency, highlighting efforts to reduce energy consumption and emissions.
    \item \textit{Energy storage} includes videos focused on technologies and strategies for storing energy, featuring specific terminology and more technical discussions.
\end{itemize}

\begin{figure}[!htbp]
\begin{center}
\includegraphics[width=\textwidth]{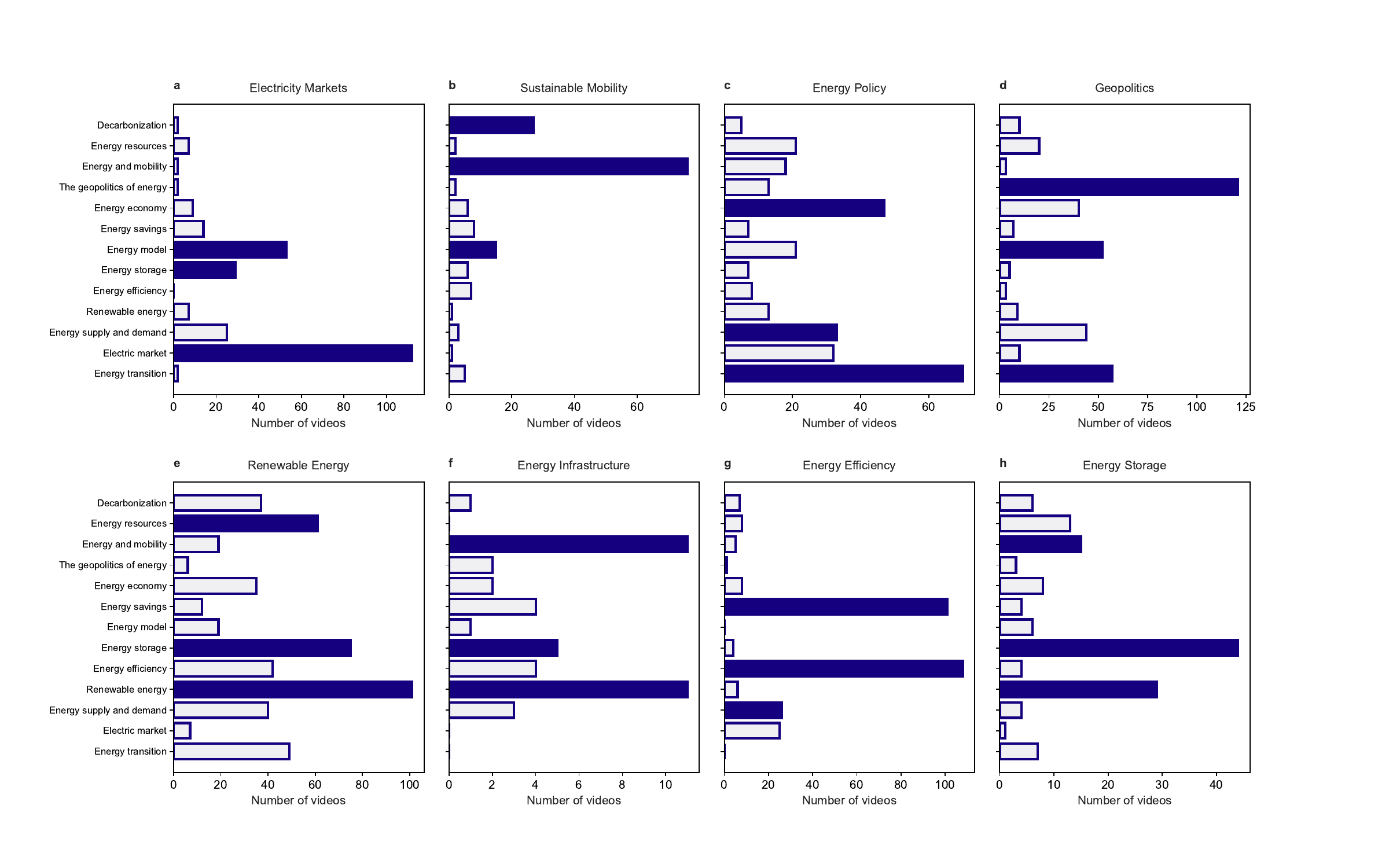}
\end{center}
\caption{\textbf{Distribution of concepts by topic.} Number of videos per concept in each topic, where the highlighted bars in blue represent values above the 75th percentile, topics in the top 25\%.}
\label{fig:topic_bykeywords}
\end{figure}

Topics with broader or more widely used vocabulary—such as \textit{renewable energy}, \textit{electricity markets}, and \textit{energy efficiency} tend to include a higher number of videos. In contrast, topics with more specialised or technical focus—such as \textit{energy infrastructures}, \textit{sustainable mobility} (including discussion on electric vehicles), and \textit{energy storage} generally correspond to a smaller number of videos (Fig. \ref{fig:topic_bykeywords}).

\begin{figure}[!htbp]
\begin{center}
\includegraphics[width=\textwidth]{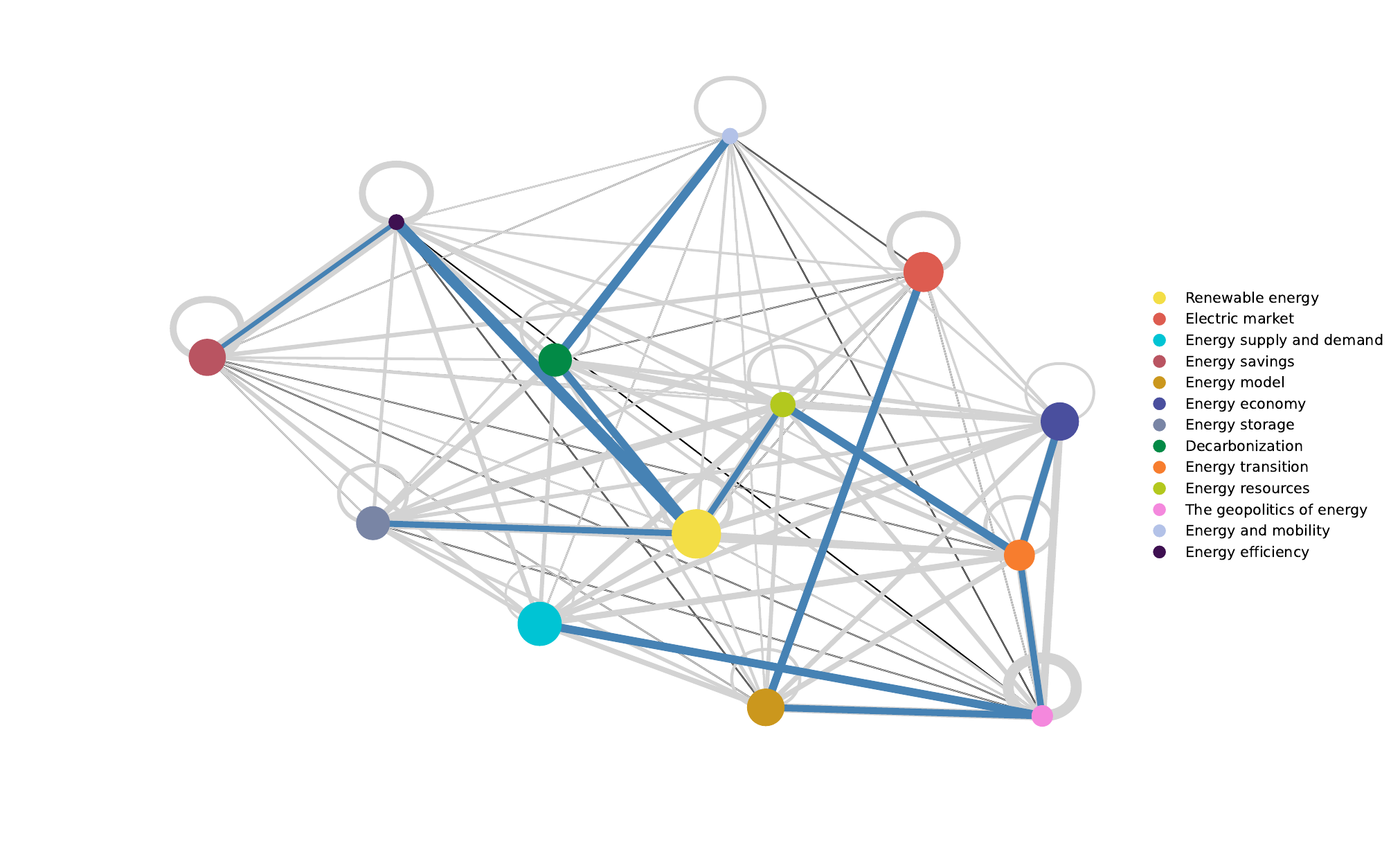}
\end{center}
\caption{\textbf{Concept network.} Network between concepts based on the number of videos that share the attributed topic. The thickness of the links is determined by the number of concept videos assigned to the same topic, normalised by the number of possible pairs of videos from the two connected concepts. The links in blue correspond to the minimum spanning tree and the dot size to the concept outflow.}
\label{fig:topic_network}
\end{figure}

We constructed a network in which nodes represent the 13 original concepts, and links reflect videos from two concepts assigned to the same topic. To account for differences in concepts' frequency, each link weight is normalised by the number of videos associated with the connected concepts. We show the structure of this network in Fig. \ref{fig:topic_network}, which illustrates the relations between concepts. We have assessed the most connected pairs and triads of keywords by calculating the weighted keyword co-occurrence given by 
\begin{equation}
W_{ij} = \sum_{t \in T} w_{ti} \cdot w_{tj}
\end{equation}
for keyword pairs, and 
\begin{equation}
W_{ijk} = \sum_{t \in T} w_{ti} \cdot w_{tj} \cdot w_{tk}
\end{equation}
for keyword triads. Where $w_{ti}$ is the number of times that keyword $i$ appears in topic $t$. The most connected pairs are: (i) the \textit{energy efficiency} and \textit{energy savings} ($W_{ij}=11577$); and (ii) the \textit{renewable energy} and \textit{energy storage} ($W_{ij}=9275$). The results suggest that videos on renewable energy are closely related to energy storage, as are the videos related to decreasing energy consumption. The most connected unique triads are: (i) the \textit{renewable energy}, the \textit{energy storage}, and the \textit{energy resources} ($W_{ijk}=481188$); and (ii) the \textit{energy transition}, the \textit{geopolitics of energy}, and the \textit{energy model} ($W_{ijk}=364356$). The concepts most isolated from the rest, computed as the ratio between the outflow (sum of outgoing weights excluding self-loops) and the total flow (sum of outgoing weights including self-loops), are the \textit{geopolitics of energy} (with 0.83) and the \textit{energy savings} (with 0.87). We have also computed the Minimum Spanning Tree (MST) to identify the connection backbone between topics since the network is fully connected. The concepts with highest betweenness centrality $b_c$ in the MST (blue edges in Fig. \ref{fig:topic_network}) are the \textit{renewable energy} ($b_c=0.65$), the \textit{energy transition} ($b_c=0.59$) and the \textit{energy resources} ($b_c=0.55$).

After extracting themes from our corpus of video transcripts, we conducted sentiment and content analysis to detect and quantify relevant semantic features. We analysed the polarity of the transcripts and classifications related to emotions, morality, semantic areas, and formality. We used the VADER model to quantify the polarity of the transcripts and compare the emotions conveyed.
\begin{figure}[!htbp]
\begin{center}
\includegraphics[width=\textwidth]{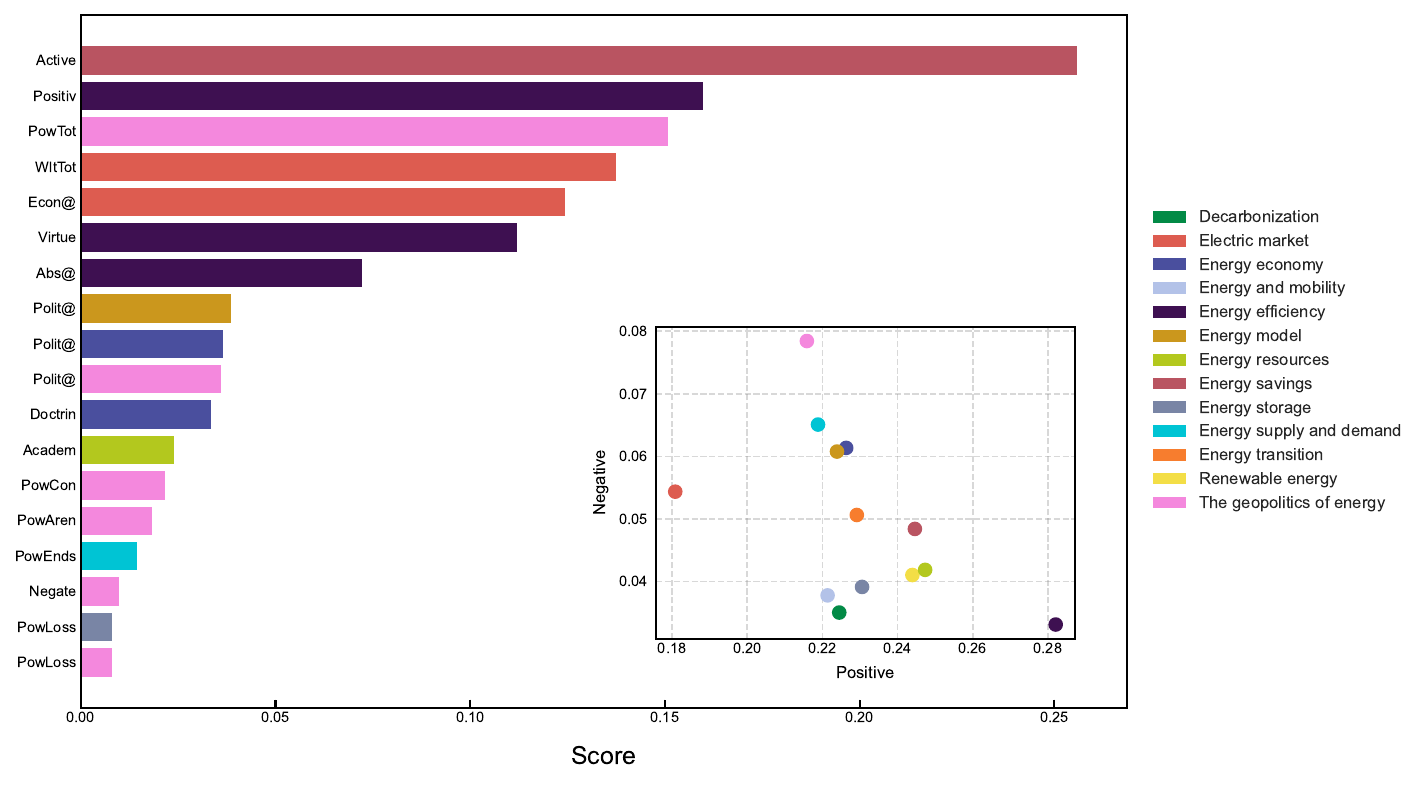}
\end{center}
\caption{\textbf{Analysis of the polar sentiment and semantic areas of the transcripts.} Concept with the highest value for each semantic area. Semantic areas are shown in the vertical axes, and the colour indicates the keyword with the highest value. Only semantic areas where a keyword has a significantly larger value than the rest are shown. The inset shows the average negative emotions as a function of positive emotions for each concept.}
\label{fig:sentiment_summary}
\end{figure}

Neutral sentiment predominates with values around 0.7 for a total of 1, implying that the discussion is carried out mostly neutrally without emotional connotations. Overall, positive sentiment exceeds negative sentiment, which only reaches 0.1 in one concept. We analysed the negative sentiment versus the positive sentiment in the Fig. \ref{fig:sentiment_summary} inset. \textit{Energy geopolitics} is the topic with the highest negative sentiment, followed by \textit{energy supply and demand}, which suggests that the perception of the political and economic situation related to energy management is the most negative among the concepts studied. The main conclusion is that concepts assigned to topics with a higher presence of economic and political terms are more negative compared to those related to \textit{energy efficiency} and \textit{energy savings}, which have a more pronounced positive connotation. The statistical tests (Fig. \ref{fig:statistical_positive} and Fig. \ref{fig:statistical_negative}) confirmed that positive and negative polarity are significantly higher in the concepts \textit{energy efficiency} (p-value  $< 4.4\cdot10^{-7}$)and \textit{energy geopolitics} (p-value $< 4.2\cdot10^{-4}$), respectively.

We further assess the semantic areas dominating each concept by calculating the average scores for the General Inquirer of the videos (the definitions of each semantic area are available in Appendix \ref{Appendix_S1}). Fig. \ref{fig:sentiment_summary} presents the concepts with significantly higher scores compared to the others (detailed values are available in Fig. \ref{fig:harvard_polar}). The \textit{energy efficiency} is significantly higher in the positive and virtue categories (p-value $< 1.22\cdot10^{-3}$), the latter defined by words indicating moral approval. The politics domain is dominated by \textit{energy geopolitics}, \textit{energy economy} and \textit{energy model}(p-value $< 3.66\cdot10^{-3}$). In addition, the \textit{energy geopolitics} is also significantly higher in the negative categories (p-value $< 3.07\cdot10^{-2}$) and most of the categories related to power, such as power control and total (p-value $< 3.16\cdot10^{-2}$). The concept of \textit{energy storage} has high power loss scores, which could imply  efficiency discussions. The \textit{electric market} displays significant higher values in the economic (p-value $< 2.06\cdot10^{-9}$) and wealth (p-value $< 8.45\cdot10^{-3}$) domains and the \textit{energy supply and demand} in the power ends  (p-value $< 6.80\cdot10^{-10}$).

\begin{figure}[!htbp]
\begin{center}
\includegraphics[width=\textwidth]{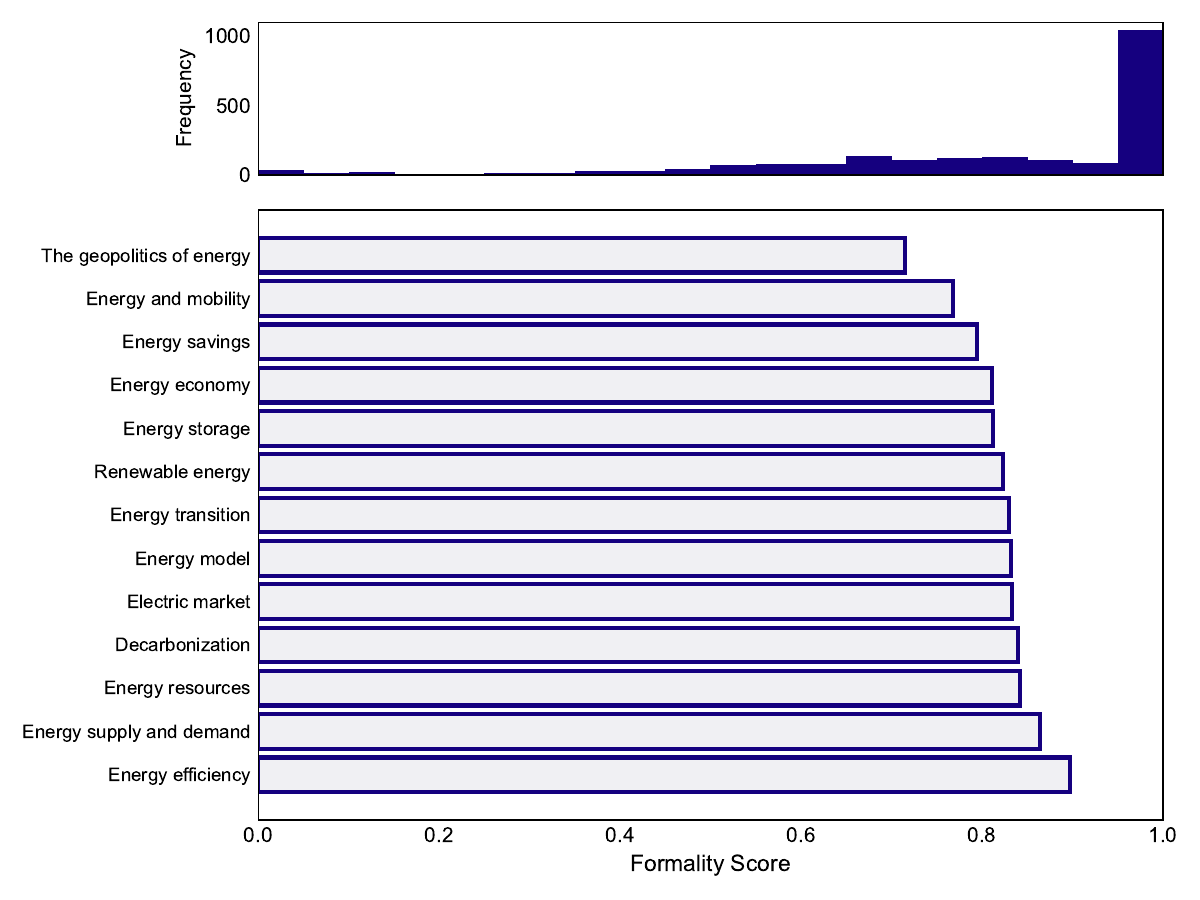}
\end{center}
\caption{\textbf{Language formality.} (\textbf{a}) Distribution of formality scores in the analysed data set. (\textbf{b}) Average formality of the videos for each of the concepts.}
\label{fig:formal_summary}
\end{figure}

An interesting feature to explore further in science communication, but not only, is the formality of language and its effect on engagement. We have computed the formality scores across our dataset, obtaining a distribution skewed to high values (Fig. \ref{fig:formal_summary}). The results suggest that the dissemination of content related to the energy transition field is overwhelmingly formal, which could indicate that it is often discussed with rigour. The average formality split by concept ranges from 0.7 to 1. While the \textit{energy efficiency} and other related subjects have the highest formality scores, the most informal concept is the \textit{energy geopolitics}.

Finally, we analysed how the different language features identified in this study correlate with user engagement on YouTube, focusing on the number of views, likes, and comments. Overall, the correlations between engagement and content features are low, with only a few reaching statistical significance. The strongest correlations (in absolute value) are found for the negative tone ($r = 0.11$, $p = 2.5 \times 10^{-7}$), loss of power ($r = 0.11$, $p = 3.9 \times 10^{-7}$), and language formality ($r = -0.11$, $p = 4.5 \times 10^{-7}$). Higher scores in negative tone and loss of power are positively associated with more comments, whereas more formal language is associated with fewer comments.

\begin{figure}[!htbp]
\begin{center}
\includegraphics[width=0.5\textwidth]{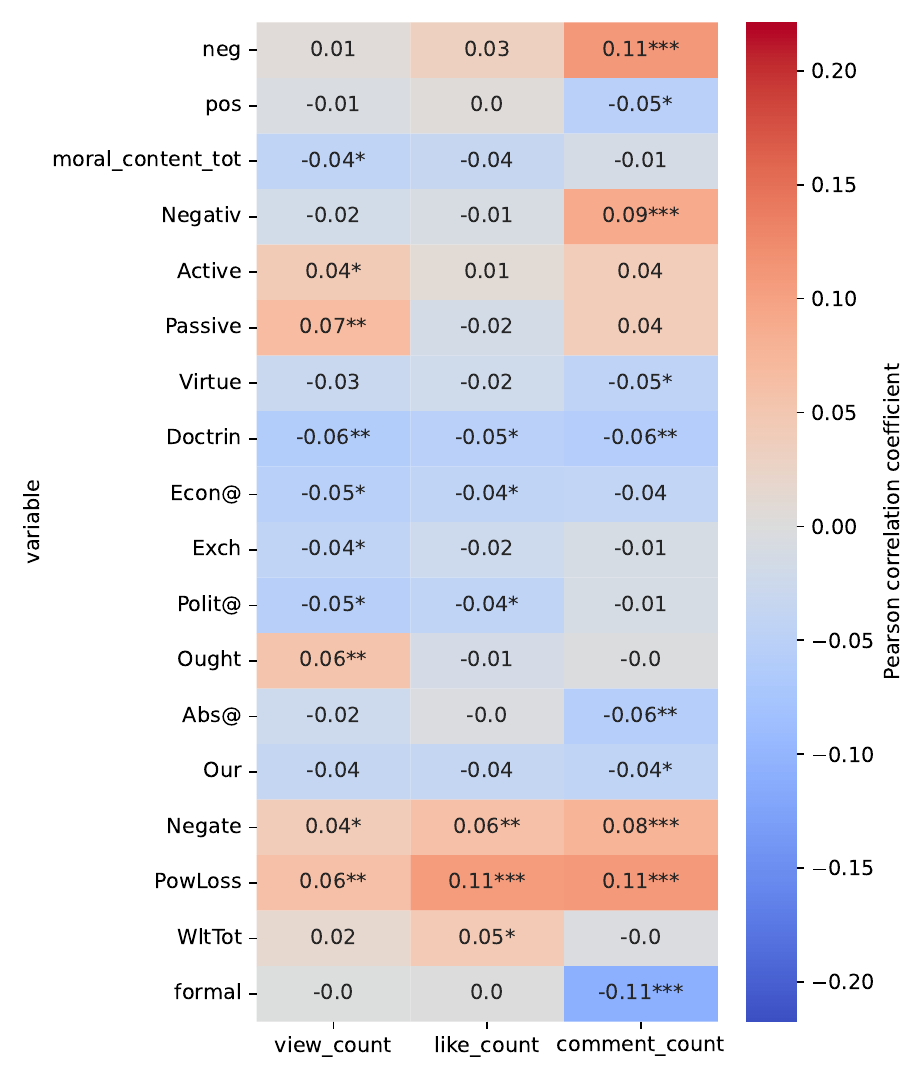}
\end{center}
\caption{\textbf{Correlation between the language features and the measures of content engagement.} Correlation of the engagement metrics (number of views, likes, and comments) with the language features. We only show the combinations in which at least one of the engagement metrics has a significant correlation. Asterisks indicate the significance of the correlations ($^{*}$p-value$<$0.05, $^{**}$p-value$<$0.01, $^{***}$p-value$<$0.001)}
\label{fig:correlations_language}
\end{figure}

\subsection*{Original content creation and diffusion}

\subsubsection*{Experimental design and content description}

We experimentally test the influence of language formality on content engagement by creating and publishing original content related to the energy transition on YouTube through two mirror channels that feature equal content but a differentiated language use. The \textit{Euro al Joule} \cite{youtubeEur2j} channel uses formal and scientific language grounded on data. It is narrated by academics, whereas the \textit{eur2j} \cite{youtubeEuroJoule} channel uses a direct style mimicking natural conversation \cite{biel2011vlogsense,burgess2018youtube}, with more casual and less technical terms, and is presented by a narrator with well-developed interpretative and communication abilities. Hereafter, we refer to \textit{Euro al Joule} as the \textit{academic channel}, and \textit{eur2j} as the \textit{conversational channel}. A view of the YouTube cover page of each channel is available in Fig. \ref{fig:channels}. We provided consistency and coherence to the experiment by designing, recording, and publishing 20 videos on 20 challenges that represent more detailed versions of the concepts previously studied (See Appendix \ref{Appendix_S2}).

The first two videos were used to assess the diffusion of the content under neutral conditions, as they were published in the channels created in the context of this research. As of February 20, 2024, the videos published in August 2024 have 18 and 31 views in the academic channel, and 29 and 24 in the conversational channel. Highlighting the difficulty of reaching a broad audience from scratch on the platform. The following three were used to conduct a series of tests with the Google Ads platform, which handles content promotion. Finally, the experiment focuses on the remaining 15 videos, whose broadcast was carried out identically on both channels. Four different content formats are published for each challenge: a promoted complete video, a non-promoted complete video for comparison, and two brief videos in standard and short versions \cite{youtubeAdviceSupport}. The details of the Google Ads campaign parameters and the content format types are available in Appendix \ref{Appendix_S2}. The duration is the major factor that distinguishes complete and brief videos, along with the absence of supporting visual information (Fig. \ref{fig:duration_distribution}). While most brief videos are under 25 seconds, the complete videos are over a minute long, without distinction between promoted and non-promoted. The complete videos on the academic channel are longer, averaging around two minutes, compared to the conversational ones, which average around one minute. We applied the same methodology as in the previous section to measure the formality of original videos. The conversational channel content has lower formality scores and a more elongated distribution towards small values. The complete videos of both channels feature lower scores, as they might have a greater vocabulary variety (Fig. \ref{fig:formality}).

\subsubsection*{Engagement analysis on YouTube}

We compared channels and formats by computing the mean and standard deviation of the three main engagement metrics: the views, likes, and comments (Table \ref{tab:video_stats}). The first general observation is that the number of likes and comments is very limited despite the promotion campaigns. There are almost no comments, and the average likes are only significant for brief videos in short format. Overall, the type of promotion has a higher effect on the content outreach than the channel and, therefore, language formality. The average number of views of the promoted complete videos is an order of magnitude higher than that of their non-promoted counterparts, going from around 10 to 100 views. Similarly, the average views of the brief content are one order of magnitude above the complete promoted one, going from around 100 to 1000 views.

\begin{table}[h]
\footnotesize

\resizebox{\textwidth}{!}{
\begin{tabular}{llrllllll}

\toprule

channel & Promoted & N & Views ($\mu$) & Views ($\sigma$) & Likes ($\mu$) & Likes ($\sigma$) & Com. ($\mu$) & Com. ($\sigma$) \\

\midrule

Academic & Comp. not prom. & 15 & 7.1 & 8.6 & 0.1 & 0.3 & 0 & 0 \\

Conversational & Comp. not prom. & 15 & 7 & 2.1 & 0.7 & 0.6 & 0.1 & 0.3 \\

Academic & Comp. prom. & 15 & 93.5 & 80.6 & 0.1 & 0.4 & 0 & 0 \\

Conversational & Comp. prom. & 15 & 100.2 & 47 & 0.2 & 0.4 & 0 & 0 \\

Academic & Brief prom. & 15 & 895.1 & 482.5 & 0.1 & 0.3 & 0 & 0 \\

Conversational & Brief prom. & 15 & 906.9 & 597.9 & 0 & 0 & 0.1 & 0.3 \\

Academic & Brief (short) prom. & 15 & 730.5 & 364.7 & 22.4 & 11 & 0 & 0 \\

Conversational & Brief (short) prom. & 15 & 936.5 & 529.5 & 21.5 & 10 & 0 & 0 \\

\bottomrule

\end{tabular}
}
\caption{Summary of video engagement statistics by channel and format type. We show the mean ($\mu$) and standard deviation ($\sigma$) of views, likes, and comments.}

\label{tab:video_stats}

\end{table}

\begin{figure}[!htbp]

  \begin{center}

  \includegraphics[width=\textwidth]{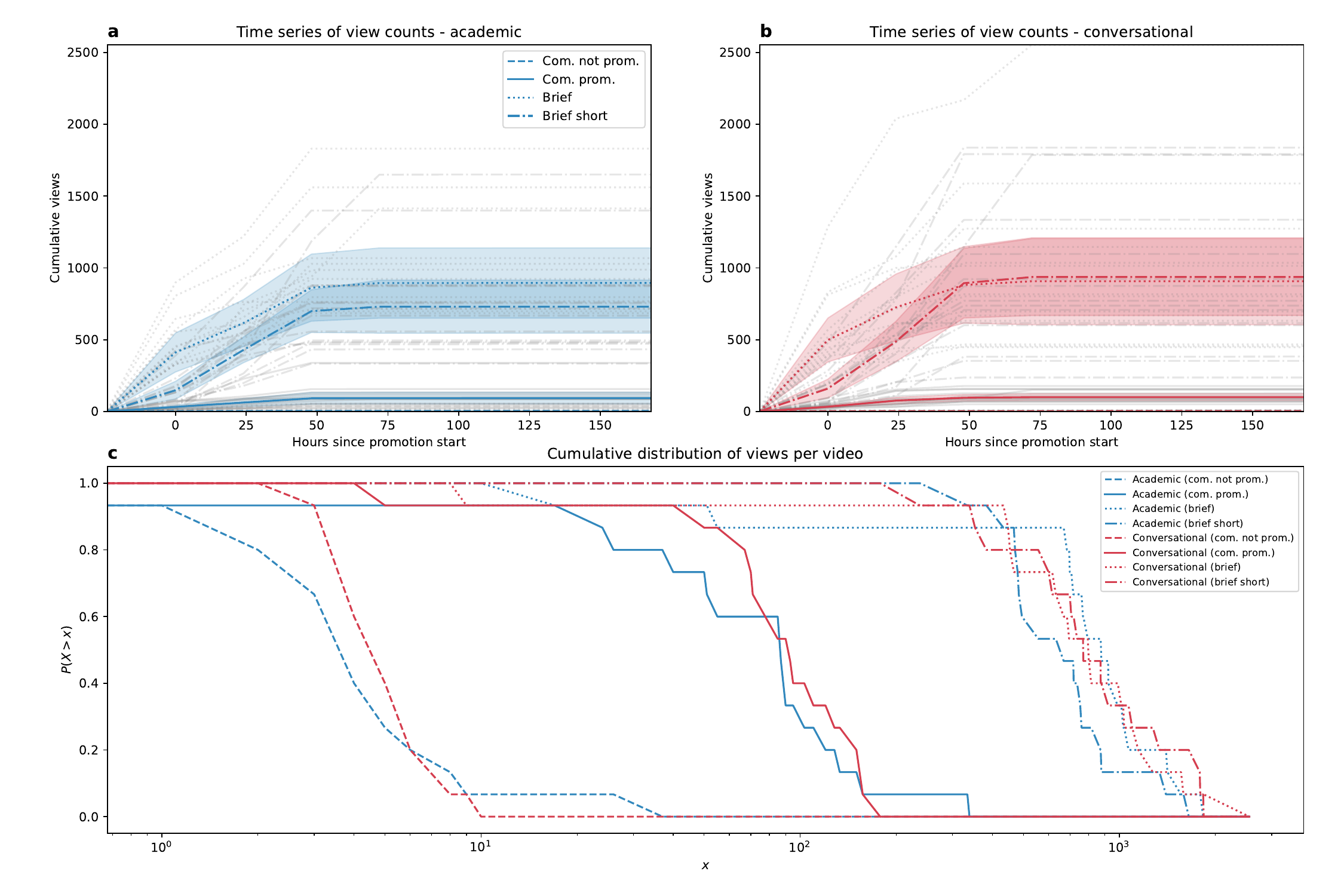}

  \end{center}

\caption{\textbf{Evolution and cumulative distribution of views.} Number of cumulative views depending on the hours since the promotion start date of each video by (\textbf{a}) the academic and (\textbf{b}) the conversational channel. We show in grey the values of each individual video, and in colour the average by type of content.(\textbf{c}) Inverse cumulative probability distributions of the views split by content type and channel.}

  \label{fig:engagement_views}

\end{figure}

We analyse the temporal evolution of views by calculating the daily cumulative views per channel and content format, aligning the content by the date of the promotion start (Fig. \ref{fig:engagement_views}\textbf{a-b}). The daily values are higher for promoted content, especially brief videos. Although conversational videos have slightly higher average views, the differences are minor compared to the effect of promotion. The maximum increase in views is reached during the two days following the start of the promotion campaign, evidencing the promotion effect. The views flatten three days after the promotion, suggesting that the content does not draw further attention. The plateau reaches similar values for both channels, with the exception of the brief videos in short format, where the academic channel has an average of 750 views, and the conversational channel of 1000.

We assess the overall differences between channels and content format by calculating the inverse cumulative probability distribution of the total views per video (Fig. \ref{fig:engagement_views}\textbf{c}). The non-promoted complete videos have the fewest views, with none of them reaching over a hundred. Promoted complete videos have around a hundred views, while brief videos in either format have around one thousand views. We did not find noticeable differences between standard brief videos and brief videos in short format. According to the distributions, the influence of the channel and language use is limited compared to the promotion. Although the conversational channel has slightly higher views, the overall differences are minor, suggesting that promotion could mitigate the role of language in content engagement and its dissemination. We have conducted the Mann-Whitney U test \cite{mcknight2010mann} to identify significant differences between the distributions (Fig. \ref{fig:views_test}). The results confirm that the differences in views per content format are statistically significant. The non-promoted content has significantly smaller views than the rest (p-value between $1.6\cdot10^{-6}$ and $3.9\cdot10^{-5}$). Similarly, the complete promoted one has significantly smaller views than the brief formats (p-value between $1.7\cdot10^{-6}$ and $2.1\cdot10^{-4}$). The comparison between the brief videos in standard and short formats did not yield significant differences in terms of views. When comparing the views between channels controlled by format, to evaluate if the language plays a role, we did not find statistically significant differences.

\begin{figure}[!htbp]

  \begin{center}

  \includegraphics[width=\textwidth]{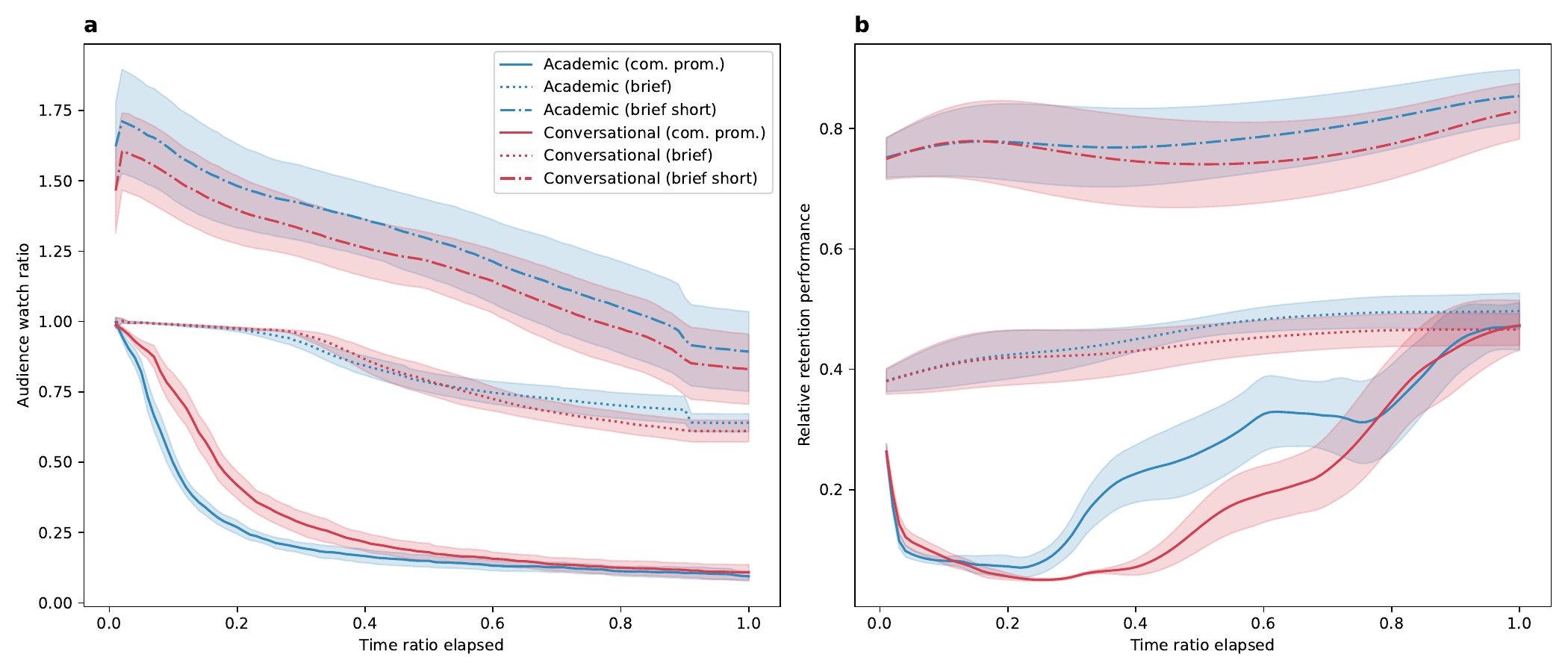}

  \end{center}

\caption{\textbf{User retention by channel and content type.} User retention metrics based on elapsed video ratio divided by content type. (\textbf{a}) \textit{audience watch ratio} and (\textbf{b}) \textit{relative retention performance}.}

  \label{fig:retention}

\end{figure}

Beyond the engagement metrics, we have also analysed user retention. Specifically, we used the \textit{audience watch ratio}, which is calculated as the number of times a part of the video was viewed divided by the total number of views, and the \textit{relative retention performance}, which captures the retention performance of a fragment compared to other videos in the platform of similar length \cite{googleMetricsYouTube}. Thus, for a \textit{relative retention performance} of 0.5, half of the videos of similar length have a better retention, and the other half have a worse retention. Similarly, a value greater than 0.5 implies that the fragment retention is better than most videos on the platform. The metrics are reported as a function of the \textit{time ratio elapsed}, which stands for the fraction of content elapsed. In Fig. \ref{fig:retention}, we show the evolution of the retention metrics split by channel and content type. The results for non-promoted complete videos are unavailable due to the limited visualisations.

According to the \textit{audience watch ratio}, almost 75\% of the audience watched the brief videos, while only 15\% watched the complete videos in their entirety. The conversational channel (eur2j) has better retention. However, its shorter duration could bias the interpretation since the \textit{time ratio elapsed} corresponds to a different temporal range in seconds. Instead, the \textit{relative retention performance} offers a more balanced view of audience retention, as it is calculated relative to content of similar length, washing out the duration effect. There are no major differences in the brief contents, but there are in the complete ones. For the complete videos, the conversational channel has a better retention in the initial fragments, whereas the academic channel has a better retention in the central fragments. To assess whether these differences are statistically significant, we calculated the two-sided Mann-Whitney U test at each time ratio (Fig. \ref{fig:retention_pvalue}), revealing that they are significant both at the beginning and in the central part. Interestingly, the channel with the highest value changes between the beginning and the middle sections. The conversational channel has a significantly higher retention between the time ratios $0.02$ and $0.07$ (p-value between $0.011$ and $0.033$) while the academic channel has a significantly higher retention between $0.23$ and $0.7$ (p-value between $0.045$ and $3.9\cdot10^{-5}$). These results suggest that conversational content is more efficient in retaining users at the beginning of the video, but academic content has better long-term retention. 

\subsubsection*{Advertisement campaign analysis}

In this section, we analyse the data from the Google Ads platform \cite{googleOverviewGoogle} where the promotion was implemented, offering a complementary view to the YouTube data. The data is reported at the campaign level, aggregating the three promoted videos by challenge, and has additional engagement metrics and audience details. We have analysed the following three metrics:

\begin{itemize}

    \item \textbf{Number of views.} Total number of views across all videos of the campaign. A view is counted when either a user interacts with the content or watches 30 seconds, or the entire video if it is shorter.

    \item \textbf{Percentage of views.} Percentage of users who watched a video when shown to them or when they saw the thumbnail. It is equivalent to the number of views divided by the number of impressions, which counts the advertisement views.

    \item \textbf{Percentage of interactions.} Percentage of users who interacted with the content when it was shown to them or when they saw the thumbnail. It is equivalent to the number of interactions divided by the number of impressions, which counts the advertisement views.

\end{itemize}

The Google Ads platform also reports details on audience characteristics. In particular, we have focused on the age and gender of users, along with the device used to watch the videos. We have analysed the audience distribution by category, finding that it is composed by a majority of men (60\%) in the older age groups (over 54) (Fig. \ref{fig:summary}).

\begin{figure}[!htbp]

  \begin{center}

  \includegraphics[width=\textwidth]{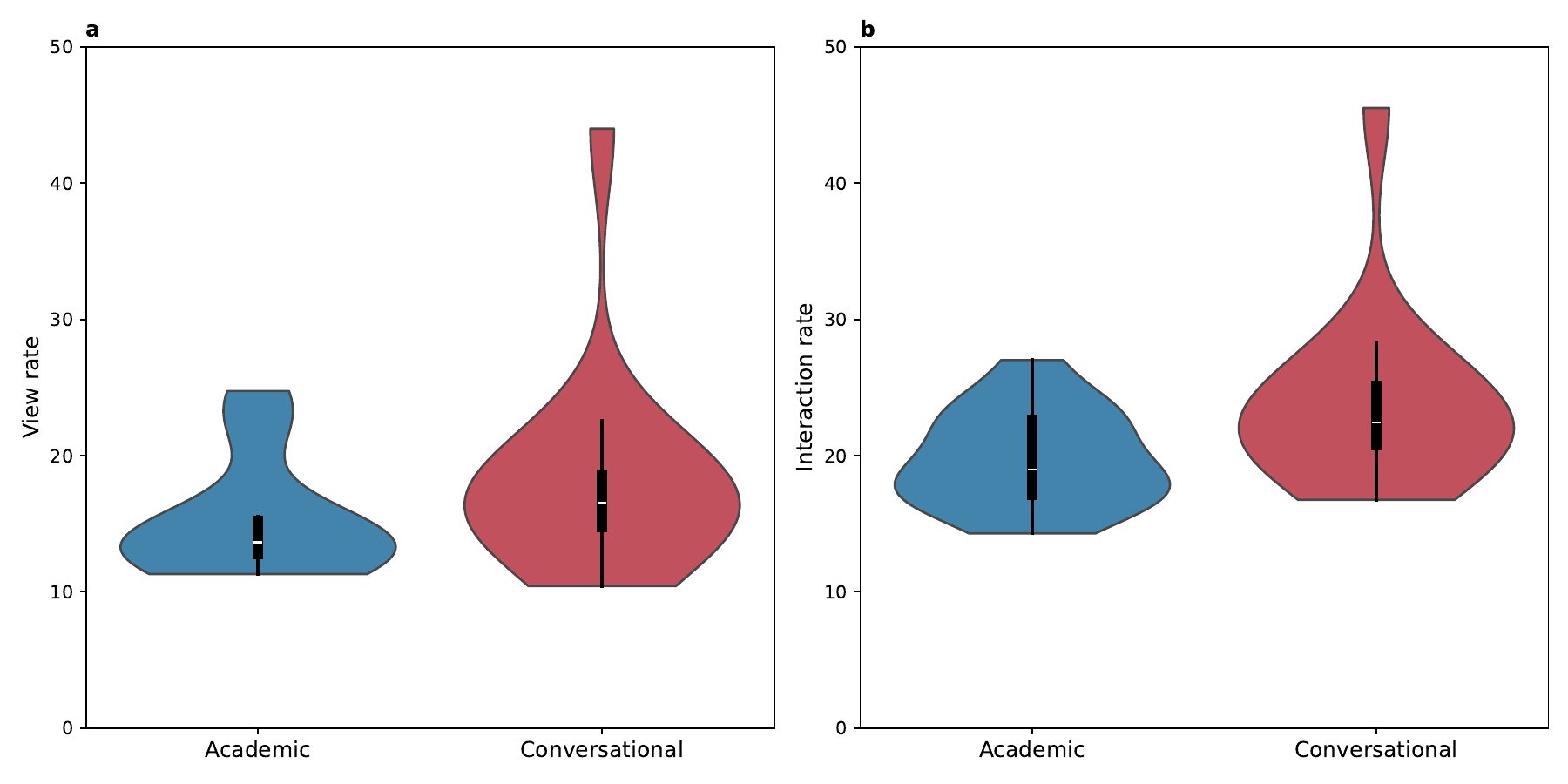}

  \end{center}

\caption{\textbf{Distribution of the view and interaction rates.} Distribution of (\textbf{a}) the viewing percentage and (\textbf{b}) the interaction rate of the promotion campaigns divided by conversational and academic content. In both cases, the distribution of conversational videos is significantly higher with p-values of $0.03$ and $0.02$, respectively.}

  \label{fig:adsgeneral}

\end{figure}

Since the number of views was already analysed in the previous section, and does not account for additional factors such as advertisement impressions, we focus here on view rates and interaction rates as more informative engagement metrics. To assess differences in engagement between channels, we compared the distributions of view and interaction percentages (Fig. \ref{fig:adsgeneral}). The results show that conversational videos exhibit significantly higher values for both metrics, indicating stronger audience engagement.

\begin{figure}[!htbp]

  \begin{center}

  \includegraphics[width=\textwidth]{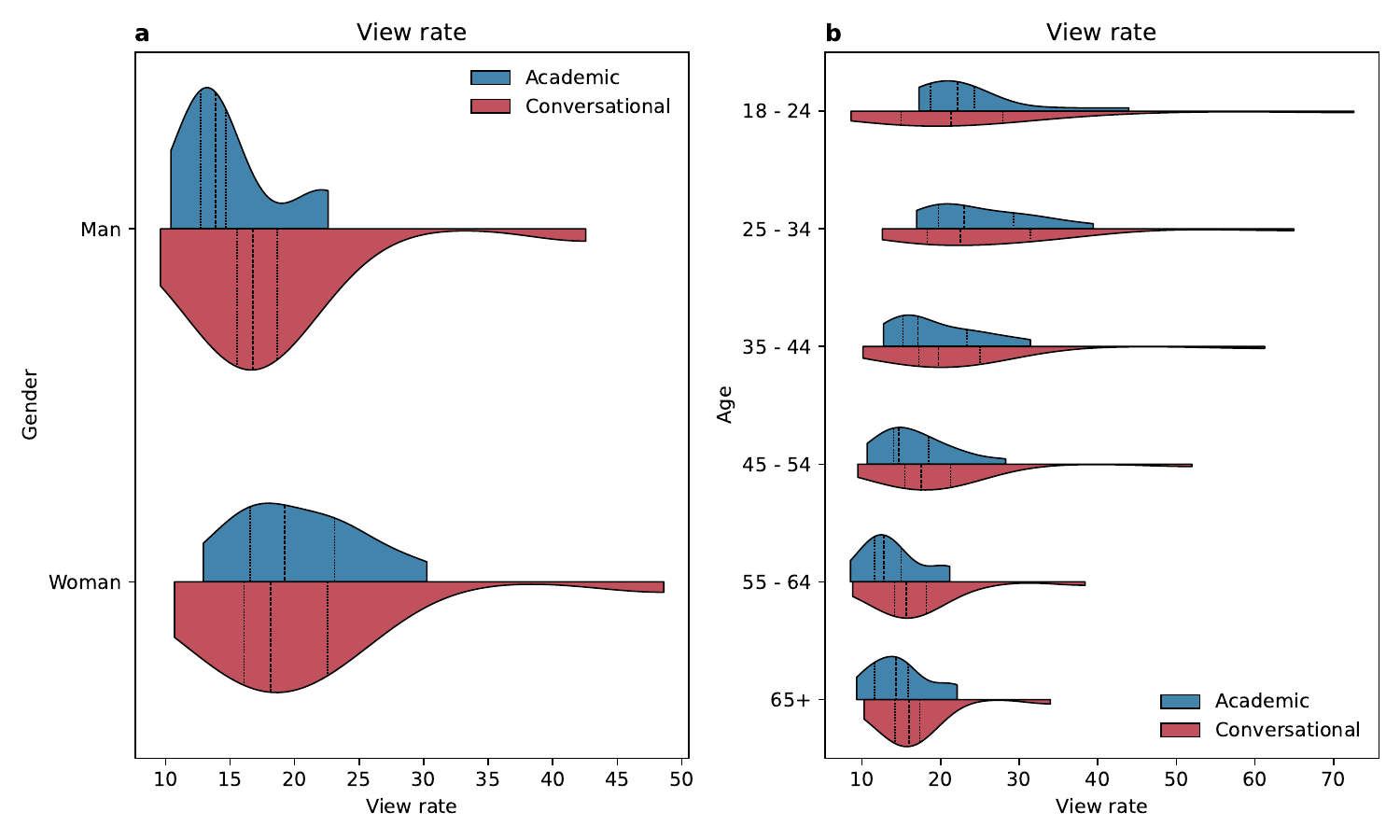}

  \end{center}

\caption{\textbf{View rate by audience profile.} Distribution of the view rate by (\textbf{a}) gender and (\textbf{b}) age.}

  \label{fig:adsdistribution}

\end{figure}

The audience information allows us to compare the engagement across demographics by plotting the view percentage distribution for each characteristic (Fig. \ref{fig:adsdistribution}). Women have higher view percentages for both channels, although the difference is more noticeable for the academic channel. Younger age groups have a higher percentage of views, especially individuals between 25 and 34. Conversational videos have a higher percentage of views across all age groups compared to the academic ones. We have conducted the Mann-Whitney U test between profiles to assess the significance of their differences (Fig. \ref{fig:adsstatiscal_views}). The main statistically significant results by gender are that the percentage of views for the academic channel among women is higher than among men, regardless of the channel (p-value of $7\cdot10^{-7}$ for the academic and $0.044$ for the conversational), but the percentages for the conversational channel between genders are not significantly different. Similarly, the view percentage of men for the academic channel is significantly lower than any other combination of gender and channel (p-value between $7\cdot10^{-7}$ and $0.018$).

The tests by age reveal that the view percentage among individuals between 25 and 34 years for both channels is significantly higher than for those aged 35 or older. Conversely, the views of academic videos by people over 55 years old are significantly lower than those under 55 years old, regardless of the channel. In agreement with the previous results, the view percentage on computers is significantly higher than for other devices. Finally, we analysed the differences between the combinations of age and gender. Women aged between 24 and 34 have a significantly higher percentage of views compared to the vast majority of other age and gender profiles. In contrast, men over 45 have a statistically significantly lower percentage of views than most profiles. In summary, while men account for a higher number of total views, young women have a higher percentage of views when we consider the impressions. We conducted the same analysis for the percentage of interactions in Fig. \ref{fig:adsdistribution_interaction} and Fig. \ref{fig:adsstatiscal_interaction}, with similar findings. The audience retention by profile is consistent with our results on the view percentage, with young individuals and women watching a larger percentage of the content (Fig. \ref{fig:adsretention}).

\section*{Discussion}

Recent events, including the Iberian Peninsula blackout or the escalation of worldwide political tensions \cite{arfaoui2025energy,kuzemko2022russia,theconversationSpainPortugalBlackouts}, have exposed the need for improved communication about the current and future challenges of energy management. This need is further amplified by the increasing risk of polarisation and political manipulation around the subject \cite{zuk2020unpacking,valquaresma2024renewable,burgess2024supply}. In the first part of this research, we analysed content from online social media platforms, motivated by the growing reliance on these channels for information consumption, particularly among younger generations \cite{dam2019engaging}. We have characterised the YouTube content related to the energy transition, focusing on the topics discussed, the associated semantic areas, the emotions conveyed, and their language formality. The analysis of YouTube content revealed that there is a wide range of subtopics discussed around the energy transition including practical subjects such as \textit{energy efficiency}, focused on improving the efficiency of buildings and houses, and more political subjects such as \textit{energy economy} or \textit{energy geopolitics}, which explore market prices and examine how nations influence and shape global energy systems through policy decisions, resource control, and alliances. The polarity around each of the concepts is similar, with high values of neutrality, evidenced by the lack of terms with strong positive or negative connotations in the topic analysis. The results suggest that the energy transition is discussed in a factual manner without strong affective polarization. Despite the emotional neutrality of the content, we identified certain outliers with higher emotional scores. Whereas the \textit{energy efficiency} content is more positive, the \textit{energy geopolitics} content is paired with negative emotions. The formality analysis aligns with the high neutrality scores as the content is shaped by formal language. Notably, the concept with higher negativity, \textit{energy geopolitics}, also features lower formality scores. The results of the correlation analysis indicate that content characterised by negative sentiment and semantic themes related to power loss tends to generate a higher number of comments. In contrast, a higher level of formality is associated with reduced audience engagement, reflected by a lower comment count. Our results agree with previous research  \cite{munaro2021engage, robertson2023negativity}, and highlight the role of emotions in audience engagement. Although studies on the effect of language formality in engagement are limited, previous results from other platforms and contexts have found a positive relation between language informality and higher user engagement \cite{rennekamp2021linguistic, wu2022promote}. 

In the second part of our work, we empirically tested the relationship between language formality and engagement. Previous studies have already raised doubts about the role of content, showing how videos with high-quality content do not engage more than their low-quality counterparts \cite{desai2013content}. Similarly, user-generated content receives a higher engagement than professional content \cite{welbourne2016science}. Our initial content publication in the newly created channels revealed difficulties in reaching a broad audience without prior visibility. To achieve a sufficient number of views for in-depth statistical analysis, we used the Google Ads platform to promote the content. The process of promotion had a significant effect compared to the base scenario, with a visualisation increase of three orders of magnitude. Although the conversational channel received slightly more views than the academic one, the difference was not statistically significant. The small sample size and high variance of engagement across videos might affect the results of the statistical tests. The detailed data reports from Google Ads enabled a fair comparison between channels and an assessment of the interest by demographic characteristics. The comparisons revealed that view and interaction rates are significantly higher in the conversational channel, suggesting that direct and casual language may improve user engagement and broaden the audience of energy transition content. The results on view and interaction rates complement previous studies that analysed a popular science YouTube channel and did not use promotion campaigns \cite{yang2022science}. While they found that the majority of the audience was male, our results suggest that women are more engaged with the content when normalised by impressions. Our work also highlights that young people are more interested in the content related to the energy transition \cite{dam2019engaging}. Notably, interest varies by age, gender, and other socioeconomic characteristics depending on the subject.

Our work summarises the discussions around the energy transition and assesses how language features can condition user engagement from an analytical and experimental perspective. Our work provides evidence on the need to reshape scientific dissemination to provide more trustful information on the energy transition and the challenges arising. Whether young individuals are inherently more interested in the content, or additional factors like platform typology or recommendation algorithms are involved, remains an open question. Our work emphasises the need for experimental studies to obtain insights into the role of content format in user engagement. The results might not be extrapolable to all social media but vary across platforms and audiences. Our static perspective opens the door to longitudinal studies that assess the temporal evolution of engagement in energy transition content based on internal platform dynamics and external events. Similarly, qualitative studies could provide more insights into the reasons behind the interest disparities in the content by sociodemographics.

\section*{Author contributions}
\begin{itemize}
    \item Conceptualization: Aleix Bassolas, Piero Birello and Julian Vicens.
    \item Data curation: Aleix Bassolas and Pierlo Birello.
    \item Formal analysis: Aleix Bassolas and Piero Birello.
    \item Investigation:  Aleix Bassolas.
    \item Methodology:  Aleix Bassolas, Piero Birello and Julian Vicens.
    \item Supervision: Julian Vicens.
    \item Writing – original draft: Aleix Bassolas.
    \item Writing – review \& editing: Piero Birello, Julian Vicens.
\end{itemize}

\section*{Acknowledgments}

We thank our project partners, particularly Daniel Campos and F. Xavier Alvarez, for their valuable comments on the paper and for their contributions in creating the audiovisual content analysed in this work. 

\section*{Funding}
This work was supported by project 23S06035-001, funded by the 2023 Research and Innovation Grant Program of the Barcelona City Council in collaboration with the “La Caixa” Foundation.

\section*{Data availability statement}

The YouTube identifiers of the content analysed in this work are available at \url{10.5281/zenodo.17660511}.

\bibliography{references}

\clearpage

\onecolumngrid

\clearpage

\setcounter{figure}{0}
\setcounter{table}{0}
\setcounter{section}{0}
\setcounter{equation}{0}

\renewcommand{\thefigure}{S\arabic{figure}}
\renewcommand{\thetable}{S\arabic{table}}
\renewcommand{\thesubsection}{S\arabic{section}}  
\renewcommand{\theequation}{S\arabic{equation}} 
\renewcommand{\thefigure}{S\arabic{figure}}
\renewcommand{\thetable}{S\arabic{table}}

\appendix

\section{Analysis of concepts related to energy}\label{Appendix_S1}

\subsection*{General statistical analysis}

\begin{figure}[!htbp]
\begin{center}
\includegraphics[width=1\textwidth]{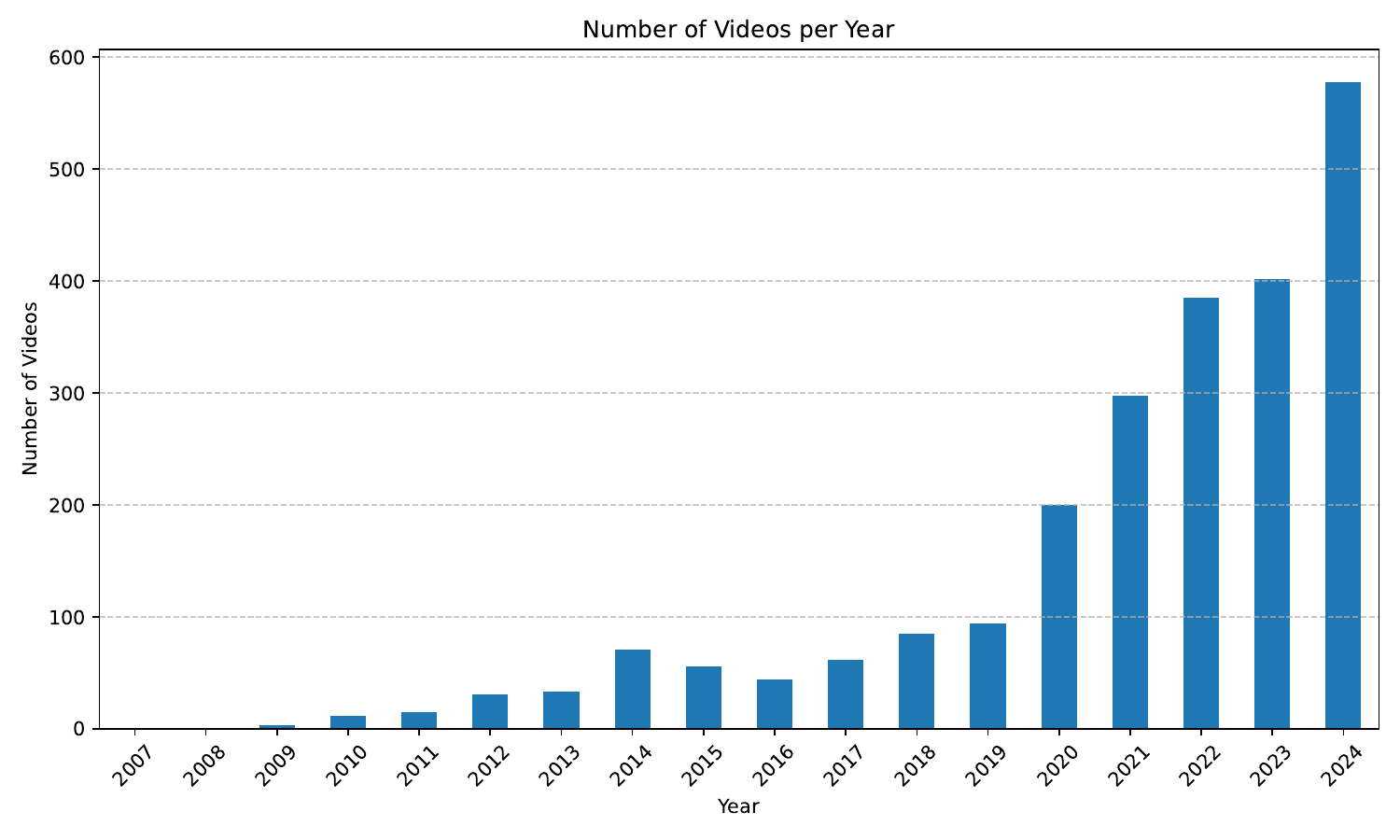}
\end{center}
\caption{\textbf{Number of videos per year.} Number of videos per year in the dataset extracted from YouTube of videos related to energy. }
\label{fig:videos_per_year}
\end{figure}

\begin{figure}[!htbp]
\begin{center}
\includegraphics[width=1\textwidth]{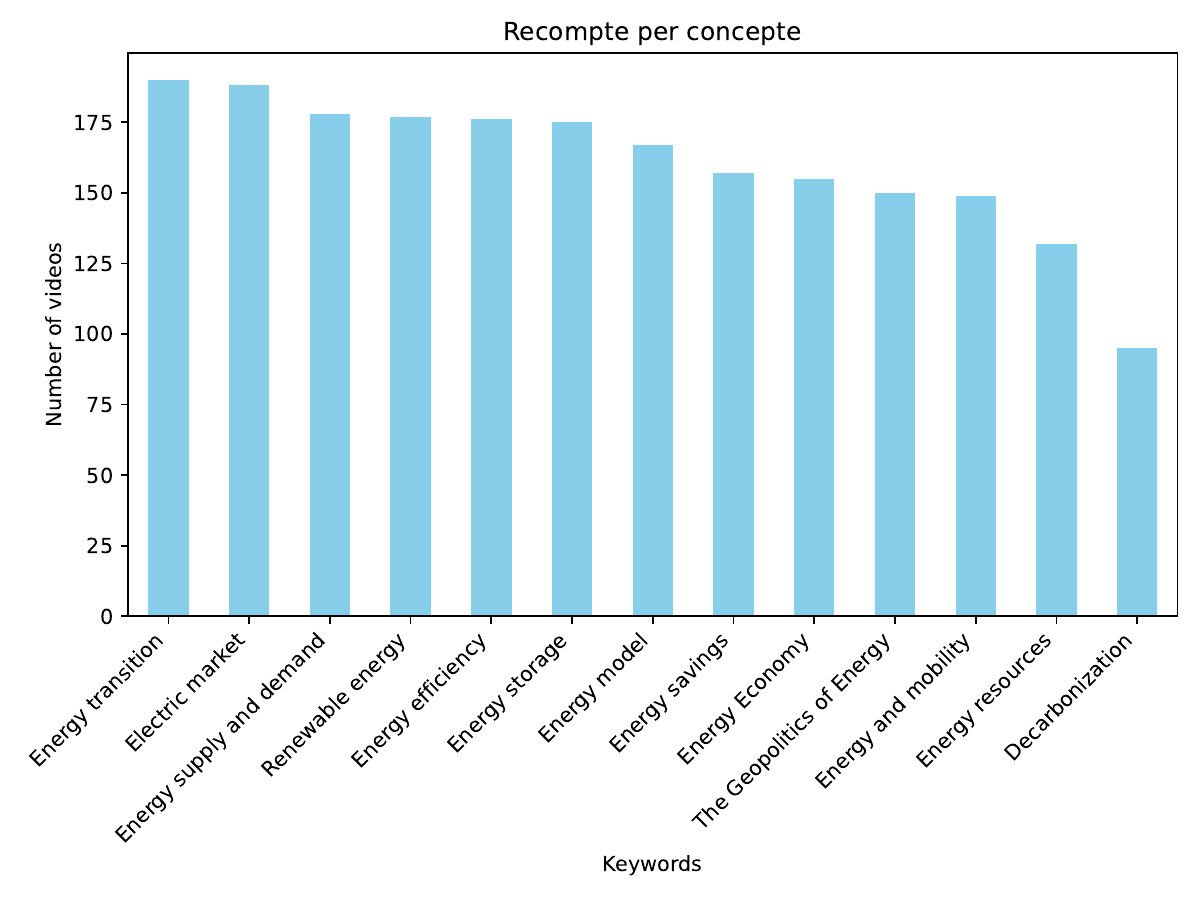}
\end{center}
\caption{\textbf{Distribution of videos per concept.} Number of videos in the dataset assigned to each concept.}
\label{fig:videos_per_concept}
\end{figure}

\begin{figure}[!htbp]
  \begin{center}
  \includegraphics[width=0.9\textwidth]{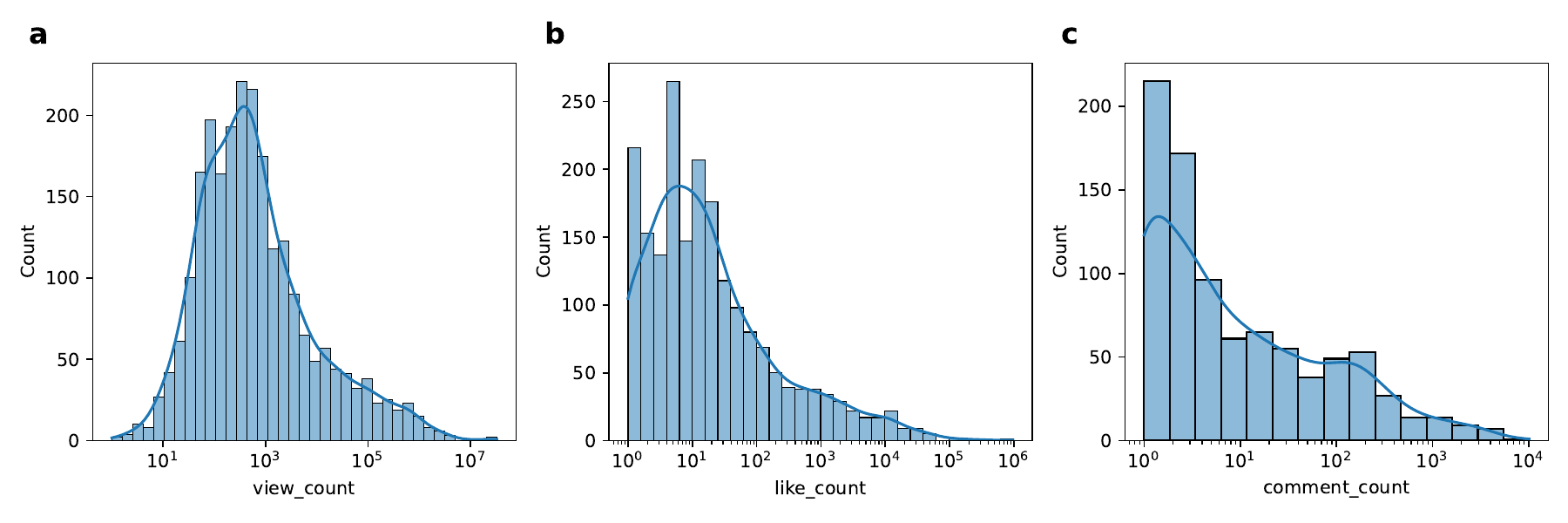}
  \end{center}
\caption{\textbf{Distribution of engagement metrics.} Distribution of \textbf{a} views, \textbf{b} likes and \textbf{c} comments.}
  \label{fig:statistics_distribution}
\end{figure}

\begin{figure}[!htbp]
\begin{center}
\includegraphics[width=\textwidth]{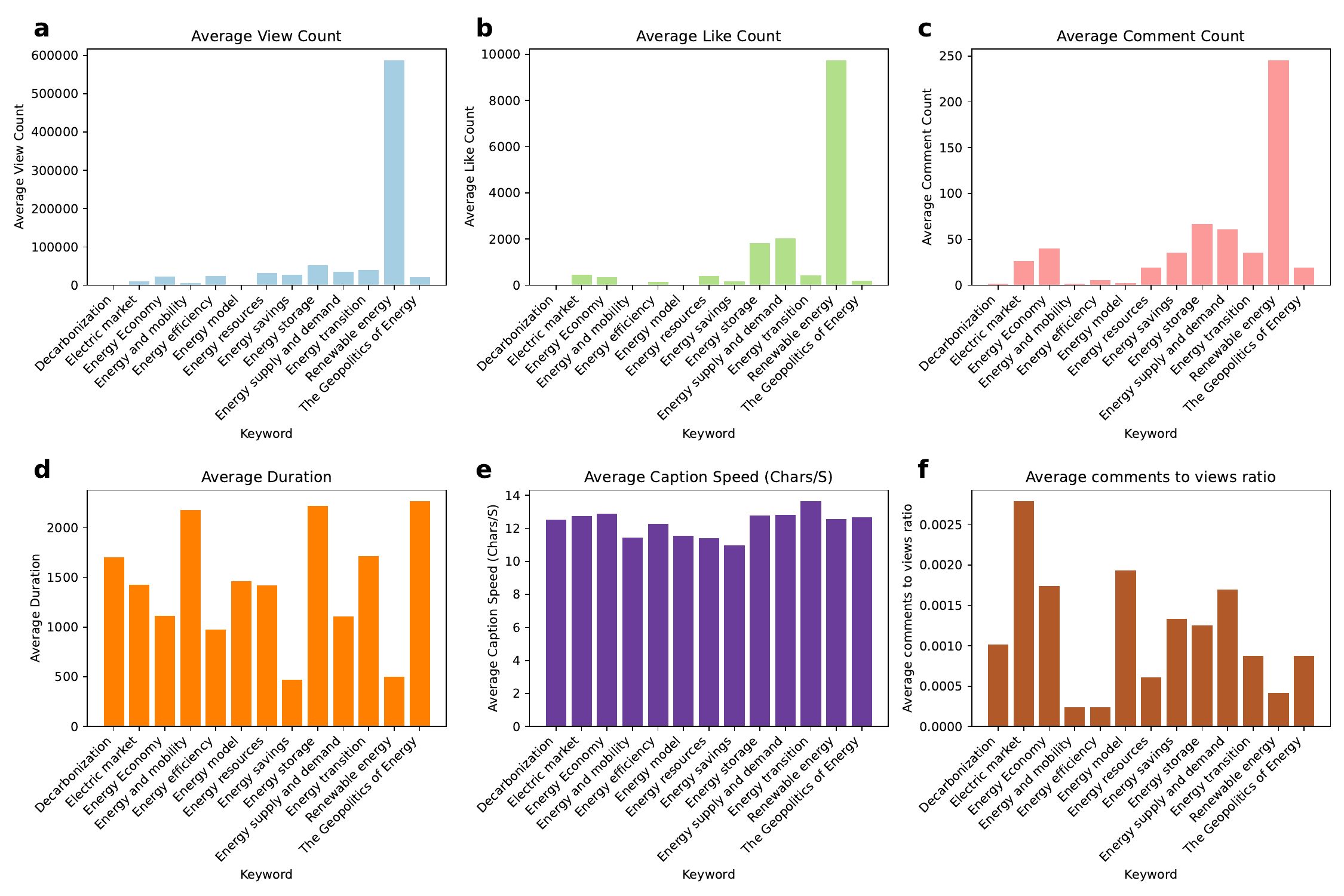}
\end{center}
\caption{\textbf{Statistics by concept for Spanish content.} Average separated by concept of the main statistics: (\textbf{a}) views, (\textbf{b}) likes, (\textbf{c}) comments, (\textbf{d}) duration, (\textbf{e}) speed of the text, and (\textbf{f}) ratio between comments and views.}
\label{fig:statistics_concept}
\end{figure}

We analyse the number of videos per year in Fig. \ref{fig:videos_per_year}, where most of them are recent, with a steeper increase after 2020. Fig. \ref{fig:videos_per_concept} has the number of videos per concept after filtering the videos in Spanish and those including terms related to energy or electricity. The concepts with more videos are the \textit{energy transition} and the \textit{electric market}. The overall distribution of the engagement metrics (views, likes, and comments) reveals that views are higher, followed by likes and comments (Fig. \ref{fig:statistics_distribution}). Despite the distributions having a clear peak, all of them are heavy-tailed, with a few videos featuring large values. In Fig. \ref{fig:statistics_concept}, we show the number of views, likes, comments, and the comment-to-view ratio. We also report two general characteristics of the content: the duration and the characters per second.  There are disparities in concept engagement, with the \textit{renewable energy} having more views, likes, and comments. Other concepts that gather more attention are the \textit{energy supply and demand} and the \textit{energy economy}. We also observe that the content has a large duration variability, ranging between 500 and 2000 seconds.  Instead, the content speed, measured as characters per second, is very stable. The content interaction, measured as the ratio of comments and views, provides a complementary perspective to the results. In this case, concepts related to the economy, such as \textit{electric market}, \textit{energy supply and demand}, and the \textit{energy economy}, are more salient. 

\subsection*{Language analysis of energy content in YouTube}

\begin{figure}[!htbp]
\begin{center}
\includegraphics[width=0.8\textwidth]{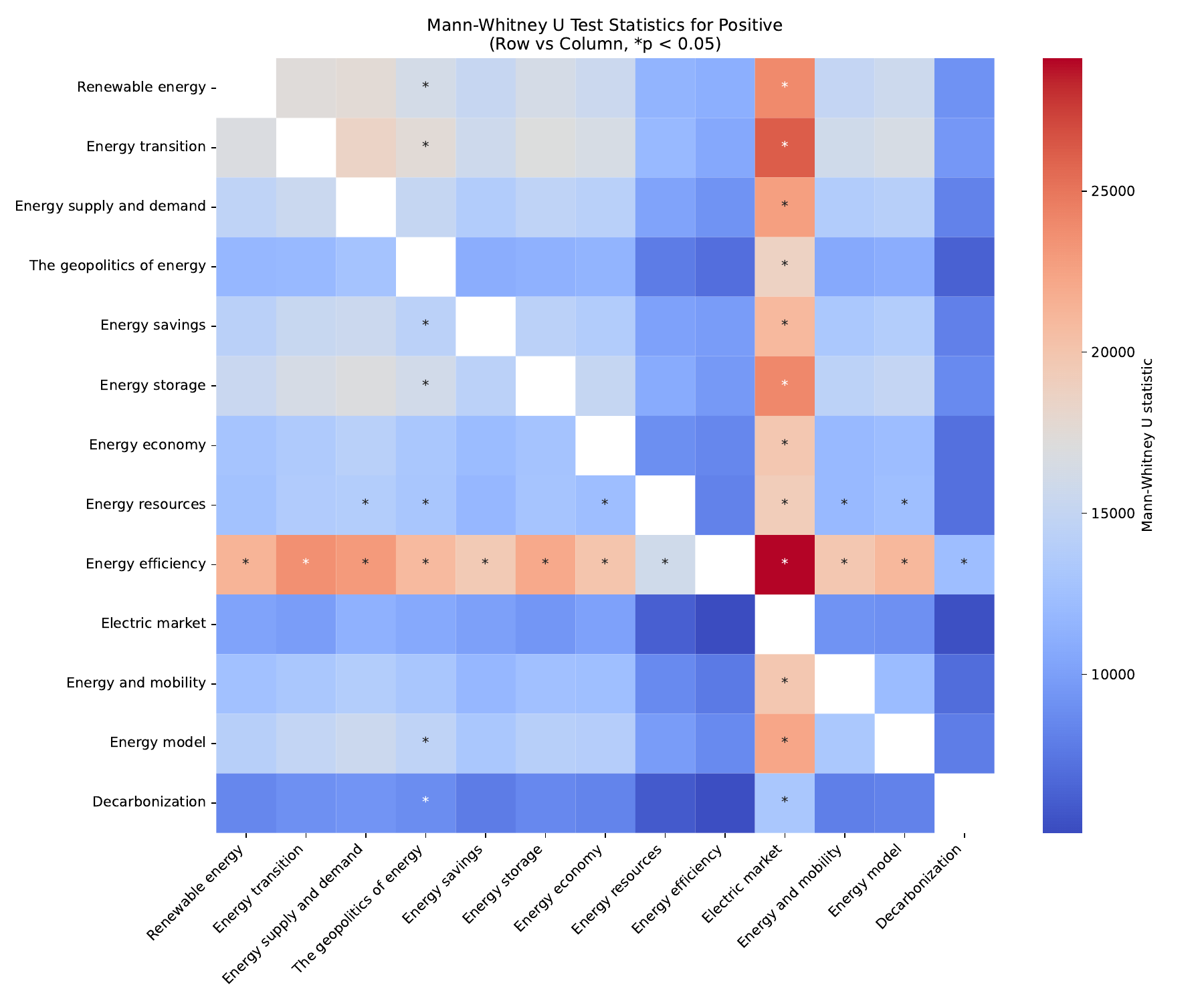}
\end{center}
\caption{\textbf{Statistical tests between positive polarity across concepts.} Mann-Whitney U test between the positive polarity of concepts. The comparison is performed to assess whether concepts on the vertical axis are greater than those on the horizontal axis. Asterisks indicate that differences are significant.}
\label{fig:statistical_positive}
\end{figure}

\begin{figure}[!htbp]
\begin{center}
\includegraphics[width=0.8\textwidth]{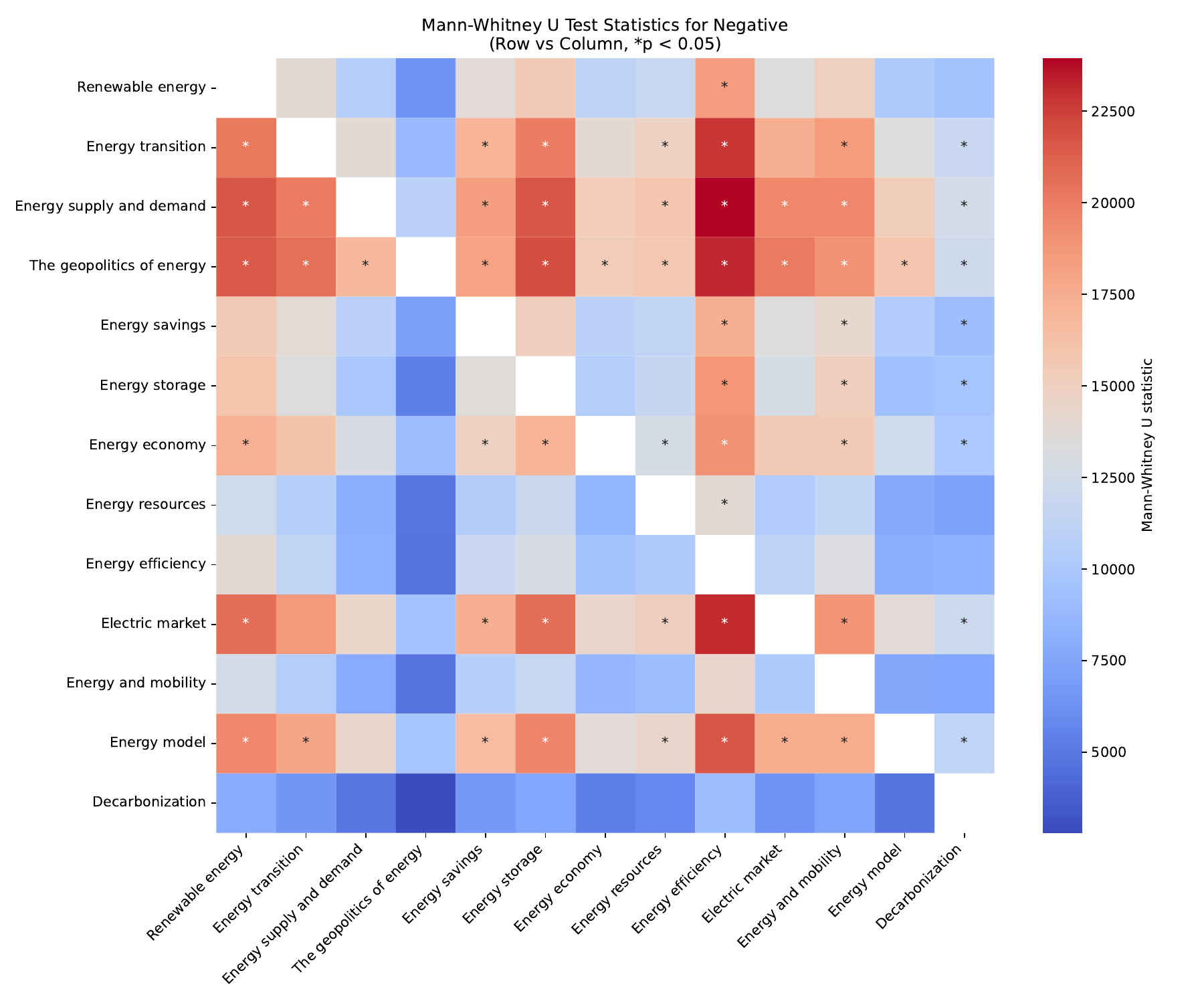}
\end{center}
\caption{\textbf{Statistical tests between negative polarity across concepts.} Mann-Whitney U test between the negative polarity of concepts. The comparison is performed to assess whether concepts on the vertical axis are greater than those on the horizontal axis. Asterisks indicate that differences are significant.}
\label{fig:statistical_negative}
\end{figure}

In Fig. \ref{fig:statistical_positive}, we provide the statistical tests between concepts for positive sentiments. The concept of \textit{energy efficiency} has significantly higher values than the rest. Instead, the \textit{electric market} has significantly lower positive scores. The other concept with significant differences in positive scores is \textit{energy resources}.
Regarding the negative sentiments (Fig. \ref{fig:statistical_negative}), multiple concepts have significantly higher values. The most notable ones are the \textit{geopolitics of energy}, the \textit{energy transition}, and the \textit{energy supply and demand}.

\subsection*{The General Inquirer}\label{subsec:methods_GI}

We report here the chosen General Inquirer categories separated into four larger classes, with their corresponding definition: 
\begin{itemize}
    \item Broad or moral categories: \begin{itemize}
        \item \textit{Positiv}: 1,915 words of positive outlook (It does not contain words for yes, which has been made a separate category of 20 entries.)

        \item \textit{Negativ}: 2,291 words of negative outlook (not including the separate category no in the sense of refusal).

        \item \textit{Active}: 2045 words implying an active orientation.

        \item \textit{Passive}: 911 words indicating a passive orientation.

        \item \textit{Virtue}: 719 words indicating an assessment of moral approval or good fortune, especially from the perspective of middle-class society.

        \item \textit{Vice}: 685 words indicating an assessment of moral disapproval or misfortune.

        \item \textit{Quan}: 314 words indicating the assessment of quantity, including the use of numbers.

        \item \textit{PowTot}: = 1,266 words for the whole domain of power.

        \item \textit{WltTot}: = 378 words in the wealth domain, where wealth is the valuing of having it. \end{itemize}

    \item Knowledge, rigor or social domains: \begin{itemize}
        \item \textit{Academ} 153 words relating to academic, intellectual, or educational matters, including the names of major fields of study.

        \item \textit{Doctrin}: 217 words referring to organized systems of belief or knowledge, including those of applied knowledge, mystical beliefs, and arts that academics study.

        \item \textit{Econ@}: 510 words of an economic, commercial, industrial, or business orientation, including roles, collectivities, acts, abstract ideas, and symbols, including references to money. Includes names of common commodities in business.

        \item \textit{Exch}: 60 words concerned with buying, selling, and trading.

        \item \textit{Legal}: 192 words relating to legal, judicial, or police matters.

        \item \textit{Polit@}: 263 words having a clear political character, including political roles, collectivities, acts, ideas, ideologies, and symbols.

        \item \textit{Causal}: 112 words denoting presumption that the occurrence of one phenomenon is necessarily preceded, accompanied, or followed by the occurrence of another.

        \item \textit{Ought}: 26 words indicating moral imperative.

        \item \textit{Abs@}: 185 words reflecting tendency to use abstract vocabulary. There is also an ABS category (276 words) used as a marker. 
        
        \item \textit{Negate}: has 217 words that refer to reversal or negation, including about 20 "dis" words, 40 "in" words, and 100 "un" words, as well as several senses of the word "no" itself; generally signals a downside view.

        \item \textit{Intrj}: has 42 words and includes exclamations as well as casual and slang references, words categorized "yes" and "no" such as "amen" or "nope", as well as other words like "damn" and "farewell".
        
        \item \textit{NUM}: 51 words indicating numbers. \end{itemize}

    \item The self and the others: \begin{itemize}
        \item \textit{Self}: 7 pronouns referring to the singular self

        \item \textit{Our}: 6 pronouns referring to the inclusive self ("we", etc.)

        \item \textit{You}: 9 pronouns indicating another person is being addressed directly. \end{itemize}

    \item Specific power categories: \begin{itemize}
            \item \textit{PowPt}: Power ordinary participants, 81 words for non-authoritative actors (such as followers) in the power process.

            \item \textit{PowCon}: Power conflict, 228 words for ways of conflicting.

            \item \textit{PowCoop}: Power cooperation, 118 words for ways of cooperating.

            \item \textit{PowAuPt}: Power authoritative participants, 134 words for individual and collective actors in the power process.

            \item \textit{PowGain}: Power Gain, 65 words about power increasing.

            \item \textit{PowLoss}: Power Loss, 109 words of power decreasing.

            \item \textit{PowEnds}: Power Ends, 30 words about the goals of the power process.

            \item \textit{PowAren}: Power Arenas, 53 words referring to political places and environments except nation-states.
    \end{itemize}
\end{itemize}

\begin{figure}[!htbp]
\begin{center}
\includegraphics[width=1\textwidth]{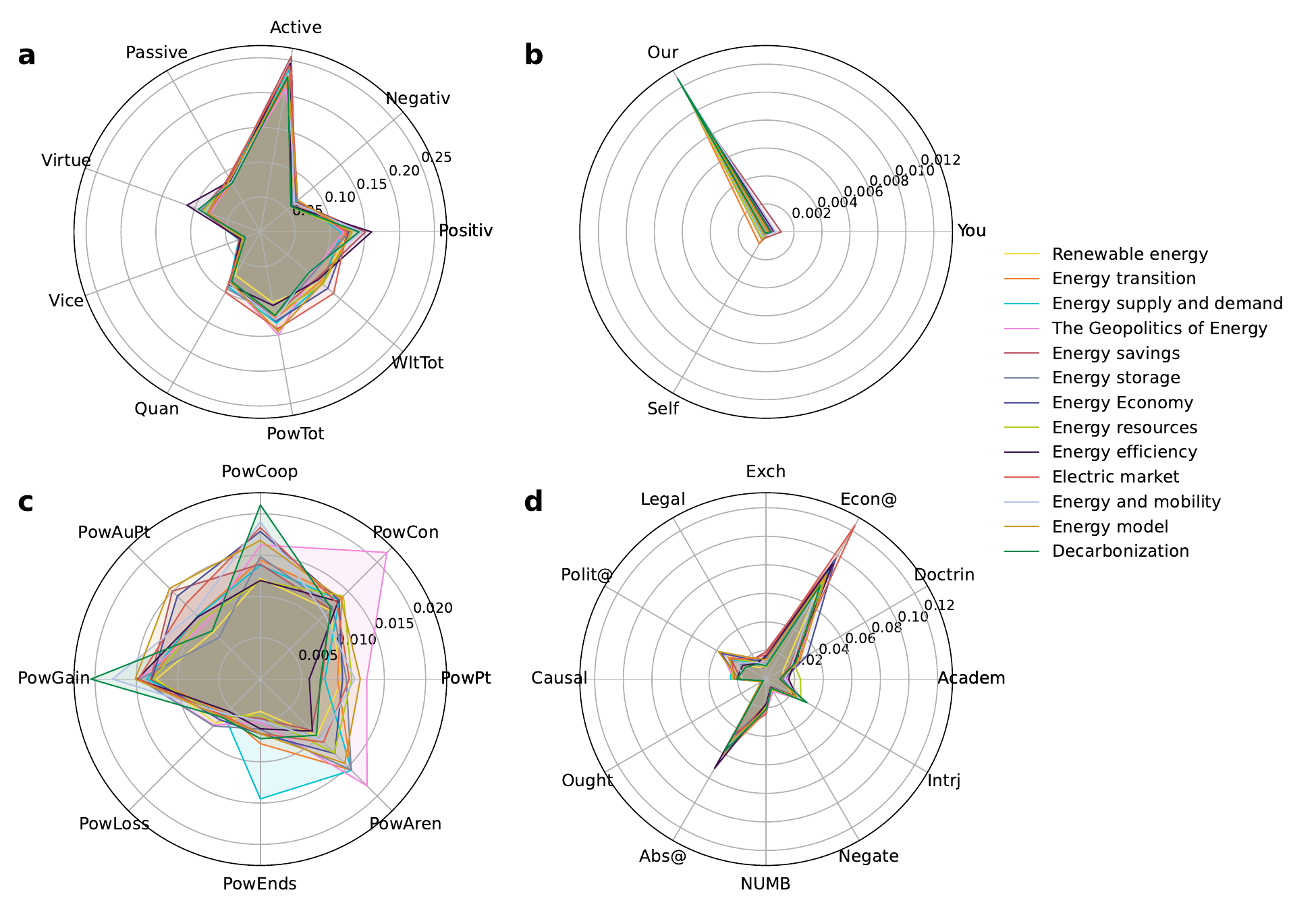}
\end{center}
\caption{\textbf{Polar plots by concept.} Scoring of the concepts by category, depending on the concept to which the content belongs. (\textbf{a}) Moral, (\textbf{b}) self and others, (\textbf{c}) power, and (\textbf{d}) knowledge categories.}
\label{fig:harvard_polar}
\end{figure}

In Fig. \ref{fig:harvard_polar}, we provide the detailed scores in each of the semantic areas analysed by concept. The \textit{energy efficiency} stands out in the values of virtue and positive domains. The \textit{electric market} has higher values in the quantitative domain and the \textit{geopolitics of energy} in the power domain. In relation to the self categories, the \textit{energy transition} has higher values in the self and the \textit{energy savings} in the you. The power domain shows a wide heterogeneity across concepts. For example, the \textit{decarbonisation} has high values in the power gain and power cooperation, the \textit{energy supply and demand} in the power ends, and the \textit{geopolitics of energy} in the power control.

\clearpage

\section{Diffusion of original content on YouTube}\label{Appendix_S2}

\subsection*{Experimental design}

\subsubsection*{Description of the challenges addressed by each video}

The 20 challenges we selected to create the YouTube videos are the following:

\begin{itemize}

\item \textbf{Challenge 1.} Determine the level of rigour of information sources in the energy field.

\item \textbf{Challenge 2.} Assess energy efficiency associated with technological improvements.

\item \textbf{Challenge 3.} Assess energy efficiency associated with behavioural changes.

\item \textbf{Challenge 4.} Identify the energy resources and technologies that govern decarbonisation strategies on a global scale.

\item \textbf{Challenge 5.} Differentiate between energy and power requirements.

\item \textbf{Challenge 6.} Distinguish the nature of the limitations associated with the use of different energy resources.

\item \textbf{Challenge 7.} Understand the statistical nature of the different future energy scenarios.

\item \textbf{Challenge 8.} Understand the implications of adopting a distributed energy model.

\item \textbf{Challenge 9.} Identify the energy needs of the different mobility models.

\item \textbf{Challenge 10.} Understand the need for the electrification of the current energy model.

\item \textbf{Challenge 11.} Understand the need for control and regulation of supply and demand in distributed energy models.

\item \textbf{Challenge 12.} Characterise the main energy storage strategies and technologies.

\item \textbf{Challenge 13.} Identify the main materials and critical flows associated with the implementation of new energy models.

\item \textbf{Challenge 14.} Critically evaluate the time scales (natural and anthropic) associated with the use of resources.

\item \textbf{Challenge 15.} Identify the implications and conflicts at the territorial management level associated with each energy model.

\item \textbf{Challenge 16.} Identify the conflicts at the level of demographic and economic management associated with each energy model.

\item \textbf{Challenge 17.} Identify the geopolitical implications of the adoption of certain energy utilisation technologies.

\item \textbf{Challenge 18.} Identify the ethical and social implications of each energy model.

\item \textbf{Challenge 19.} Assess the role of research and technological innovation in the transformation of the energy model.

\item \textbf{Challenge 20.} Understand the concepts of balance and complexity applied to the context of energy management.

\end{itemize}

\begin{figure}[!htbp]
  \begin{center}
  \includegraphics[width=\textwidth]{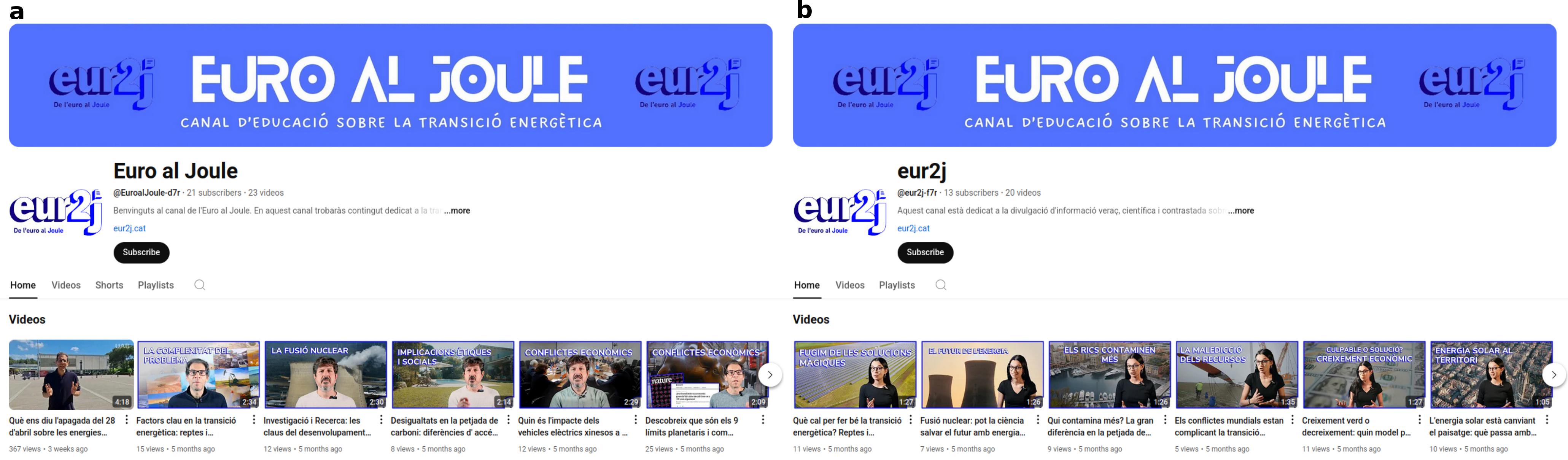}
  \end{center}
  \caption{\textbf{YouTube Cover view of each channel.} Cover page of each channel: (\textbf{a}) the academic channel (\textit{Euro al Joule}) and (\textbf{b}) the conversational channel (\textit{eur2j}.}
  \label{fig:channels}
\end{figure}

In Figure \ref{fig:channels}, we provide a view of the academic and conversational channels cover page. Their aesthetics and design are very aligned, as well as the video thumbnails differing mainly in the speaker.

\subsubsection*{Promotion campaign}

The parameters used in the promotion campaign through Google Ads are the following:

\begin{itemize}
\item \textbf{Campaign objective}. Video campaign with the view objective to achieve the maximum number of views during the promotion.

\item \textbf{Multi-format ad.} 3 versions of each video are used: long, brief, and brief in short format.

\item \textbf{Diffusion platform.} YouTube and the linked channels of the Display network. The Display Network consists of Google-owned websites such as YouTube, Google Finance, Gmail, and other addresses that serve Display advertising. It also includes a network of millions of websites and mobile apps from Google partners.

\item \textbf{User device languages.} The content is displayed for those users whose device language is set to Catalan or Spanish.

\item \textbf{Locations.} The geographical locations where the campaign has been shown are: Barcelona, Girona, Lleida, Tarragona, the Balearic Islands, and the Valencian Community.

\item \textbf{Broadcast times.} Two different times have been used during which the content is shown. \newline
From Monday to Friday: from 8:30 am to 11:45 pm. \newline
Saturday and Sunday: from 12:30 pm to 11:45 pm.

\item \textbf{Device-Specific Targeting.} The campaign is displayed on desktop computers, tablets and mobile phones. Campaign not showing on Smart TV devices.
\end{itemize}

For each channel and challenge, 4 types of video have been created that vary in length, format, and promotion. These typologies are the following:

\begin{itemize}

\item \textbf{Complete not promoted video (Complete not promoted).} This content includes the full video explaining each of the energy challenges. 

\item \textbf{Complete promoted video (Complete promoted).} The full promoted video is identical to the non-promoted one except for the opening static image.

\item \textbf{Brief promoted.} This video summarises the full video, where no supporting images are shown but only the narrator.

\item \textbf{Short video in short promoted format (Brief - short promoted).} This video has similar content to the short video, but in a short format where the image is framed vertically.

\end{itemize}

\begin{figure}[!htbp]
  \begin{center}
  \includegraphics[width=0.8\textwidth]{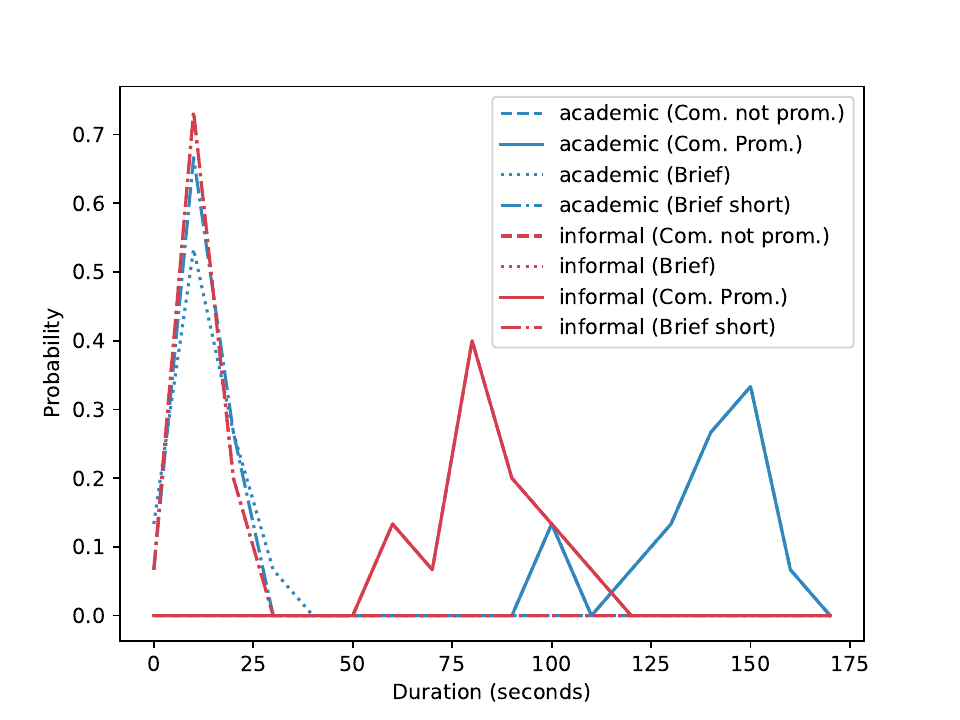}
  \end{center}
\caption{\textbf{Distribution of the duration of the videos by channel and video type.} Distribution of the duration in seconds (duration in seconds) of the videos for the Euro al Joule (blue) and eur2j (red) channels. Content has been separated by type: full videos (dashed line), full promoted videos (solid line), short videos (dotted line), and short format short videos (dashed and dotted line).}
  \label{fig:duration_distribution}
\end{figure}

Brief videos, either in short format or not, have a very similar duration across channels at around 20 seconds (Fig. \ref{fig:duration_distribution}). However, the length of complete videos vary on length depending on the channel. Whereas the video duration in the conversational channel is around one minute, the video duration in the academic channel is around two minutes.

\begin{figure}[!htbp]
  \begin{center}
  \includegraphics[width=1\textwidth]{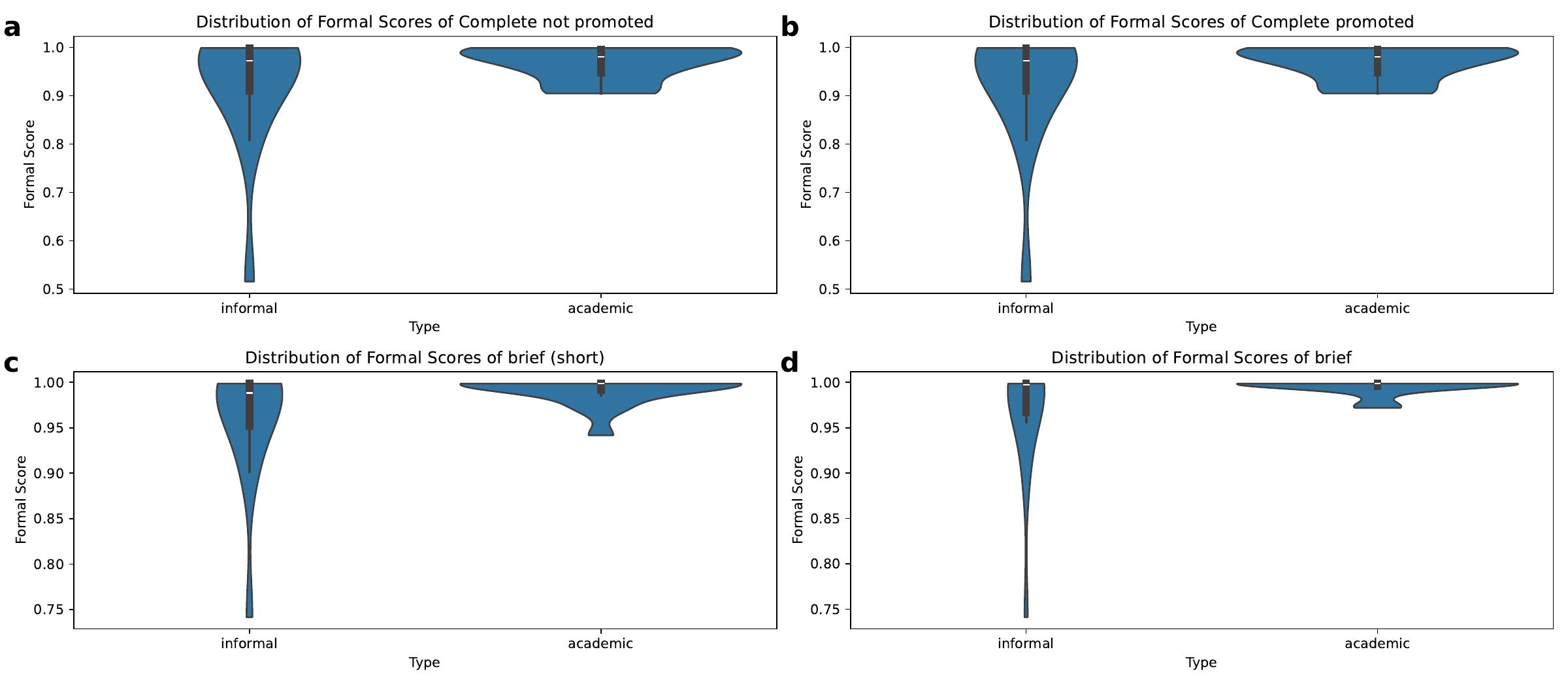}
  \end{center}
\caption{\textbf{Distribution of formality scores.} Distribution of formality scores for original videos separated by content (\textbf{a}) full non-promoted, (\textbf{b}) full promoted, (\textbf{c}) short in short format, and (\textbf{d}) standard short.}
  \label{fig:formality}
\end{figure}

We have analysed the video formality scores by channel and content format using the same methodology as in the YouTube content analysis (Fig- \ref{fig:formality}). For all content formats, the conversational channel has a lower formality average and a distribution more elongated towards lower values.

\subsection*{Analysis of YouTube data}

\begin{figure}[!htbp]
  \begin{center}
  \includegraphics[width=0.8\textwidth]{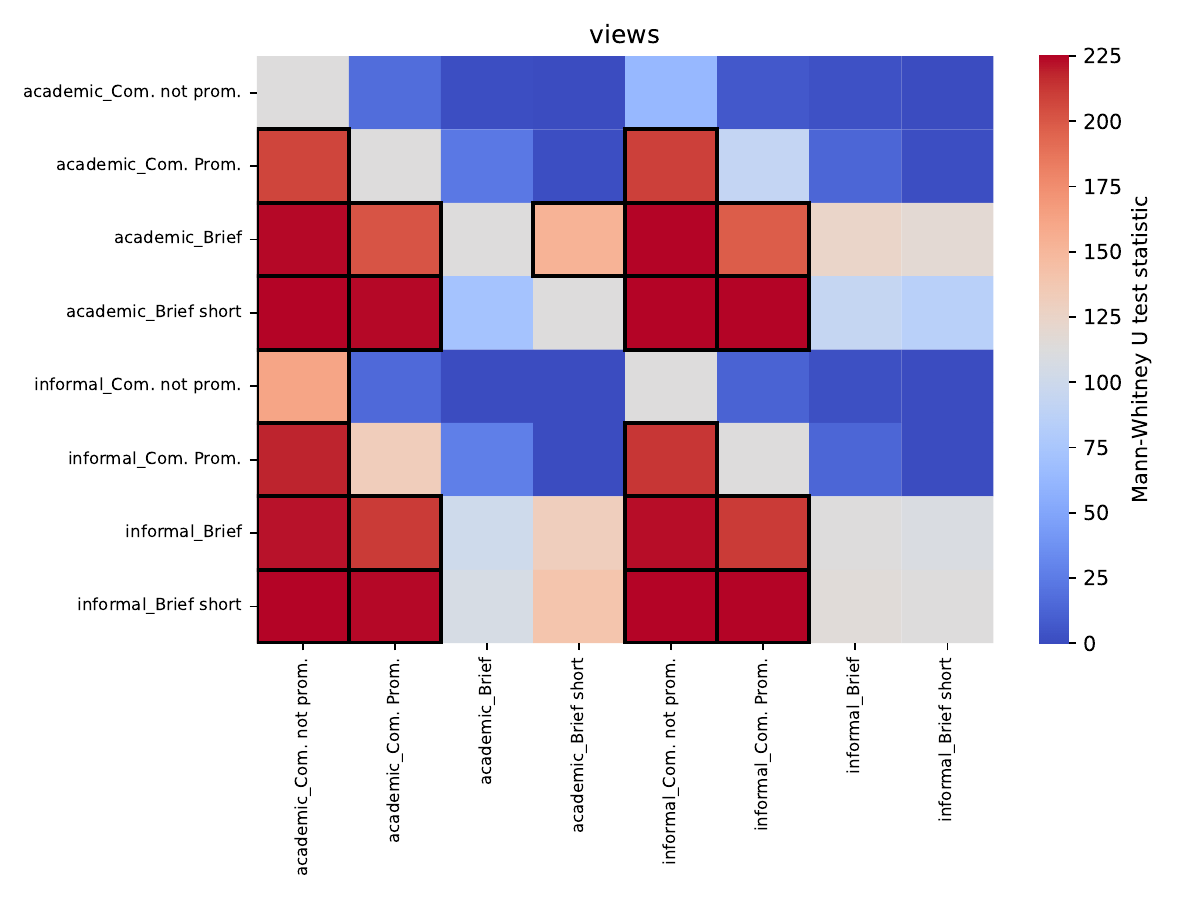}
  \end{center}
\caption{\textbf{Mann-Whitney U test between views distributions.} Results of the Mann-Whitney U test of the distributions of views per video by channel and content type. Each box includes the test value. In each statistical test, we evaluated whether the distribution of the content on the vertical axis is greater than that of the content on the horizontal axis. Entries marked in black correspond to statistically significant tests (p-value$<$0.05).}
  \label{fig:views_test}
\end{figure}

In Fig. \ref{fig:views_test}, we provide the Mann-Whitney U statistical tests between content formats divided by channel, comparing whether the entries on the vertical axes are larger than those in the horizontal axes. Entries marked in black correspond to statistically significant differences (p-value$<$0.05). There are no significant differences between channels, but there are between content format and type of promotion. Non-promoted content has significantly lower views than promoted content. Similarly, the complete promoted content has significantly lower views than brief videos. We did not find particular differences between short and standard brief videos.

In Fig. \ref{fig:views_test}, we provide the two-way Mann-Whitney U test of the relative retention performance between channels. The horizontal black indicates a p-value of $0.05$. The difference between the channels is significant during the first portion of the videos and in the central part.

\begin{figure}[!htbp]
  \begin{center}
  \includegraphics[width=0.5\textwidth]{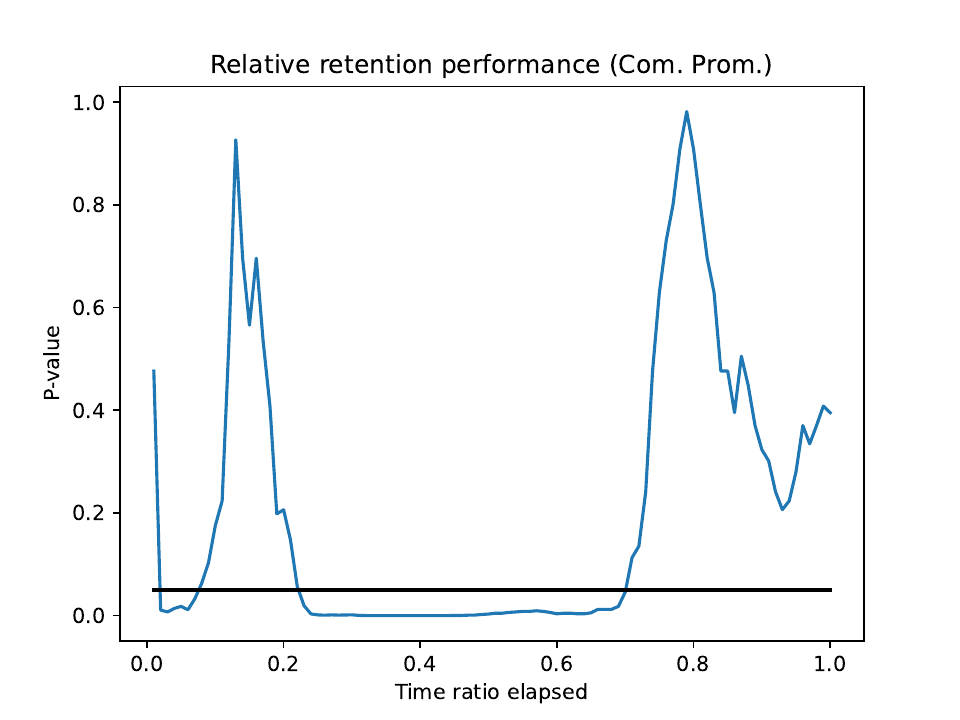}
  \end{center}
\caption{\textbf{Significance of retention values for the full promoted content.} Significance (p-value) between the relative retention performance values of each channel in the promoted long videos by elapsed video time. The black line stands for the significance value (p-value=0.05).}
  \label{fig:retention_pvalue}
\end{figure}

\subsection*{Analysis of Google Ads data}

\begin{figure}[!htbp]
  \begin{center}
  \includegraphics[width=0.9\textwidth]{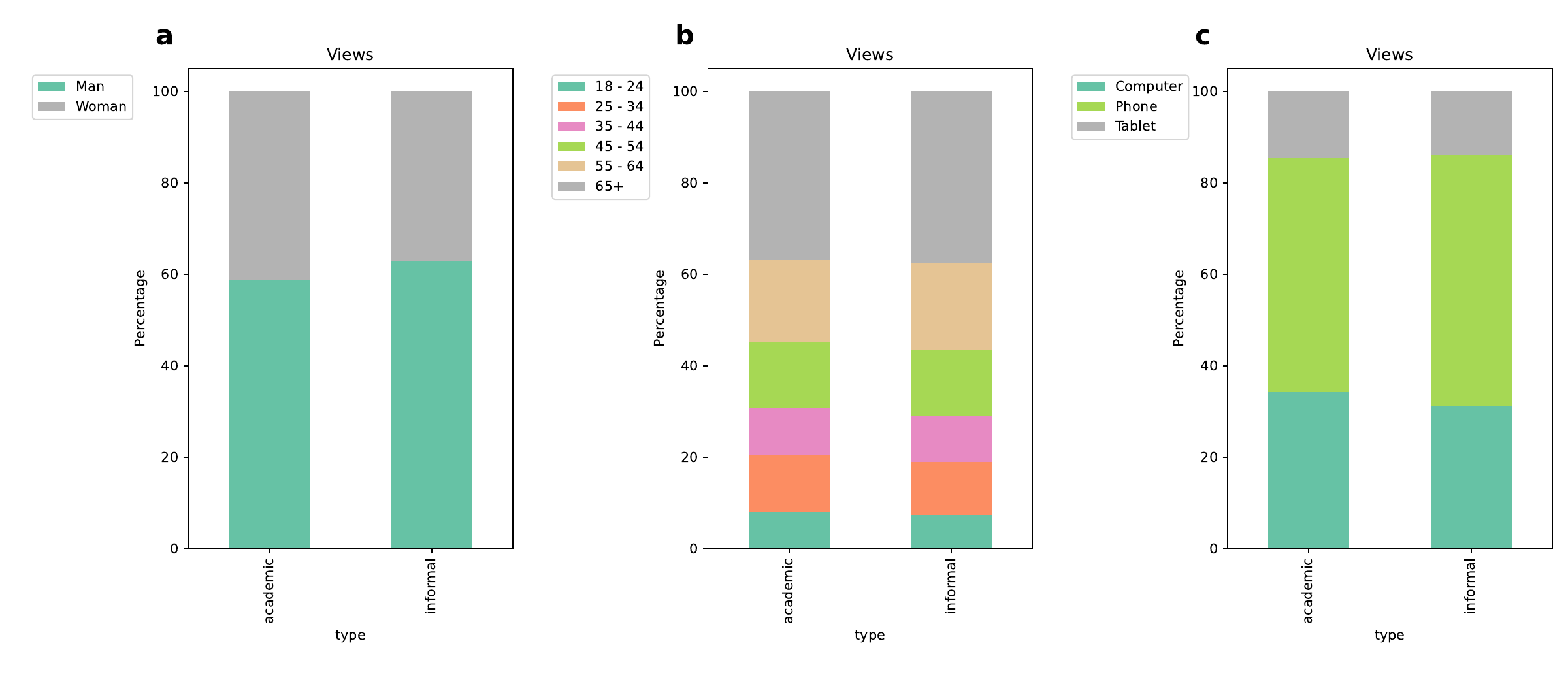}
  \end{center}
\caption{\textbf{Audience distribution} Distribution of the audience by (\textbf{a}) gender , (\textbf{b}) age (\textbf{c}) device.}
  \label{fig:summary}
\end{figure}

We have analysed the audience profile, revealing a slight over-representation of men and users older than 54 years (Fig. \ref{fig:summary}). Phones are the most common devices, although the visualisations from personal computers are not negligible.

\begin{figure}[!htbp]
  \begin{center}
  \includegraphics[width=0.9\textwidth]{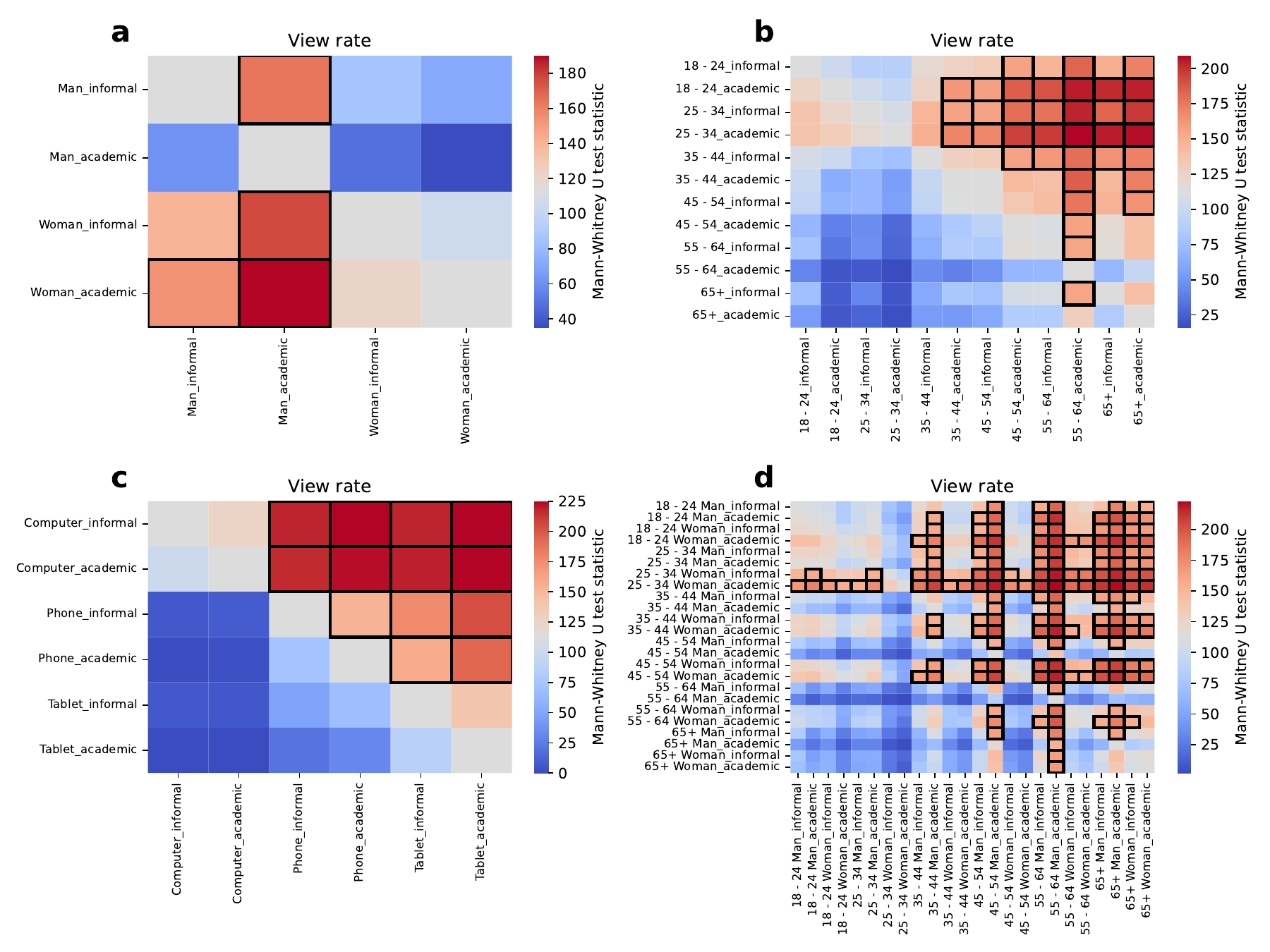}
  \end{center}
\caption{\textbf{Statistical tests between the percentage of views by audience profile.} Results of the statistical tests between the distributions of the view rate by (\textbf{a}) gender , (\textbf{b}) age (\textbf{c}) device, and (\textbf{d}) the combination of age and gender. In each statistical test, we evaluated whether the distribution of the group on the vertical axis is greater than that of the group on the horizontal axis. Entries marked in black show statistically significant tests (p-value$<$0.05).}
  \label{fig:adsstatiscal_views}
\end{figure}

We have performed Mann-Whitney U statistical tests of view rates by typology of audience (Fig. \ref{fig:adsstatiscal_views}). The high view rates from women, young users, and computers are significant in most cases.

\begin{figure}[!htbp]
  \begin{center}
  \includegraphics[width=0.9\textwidth]{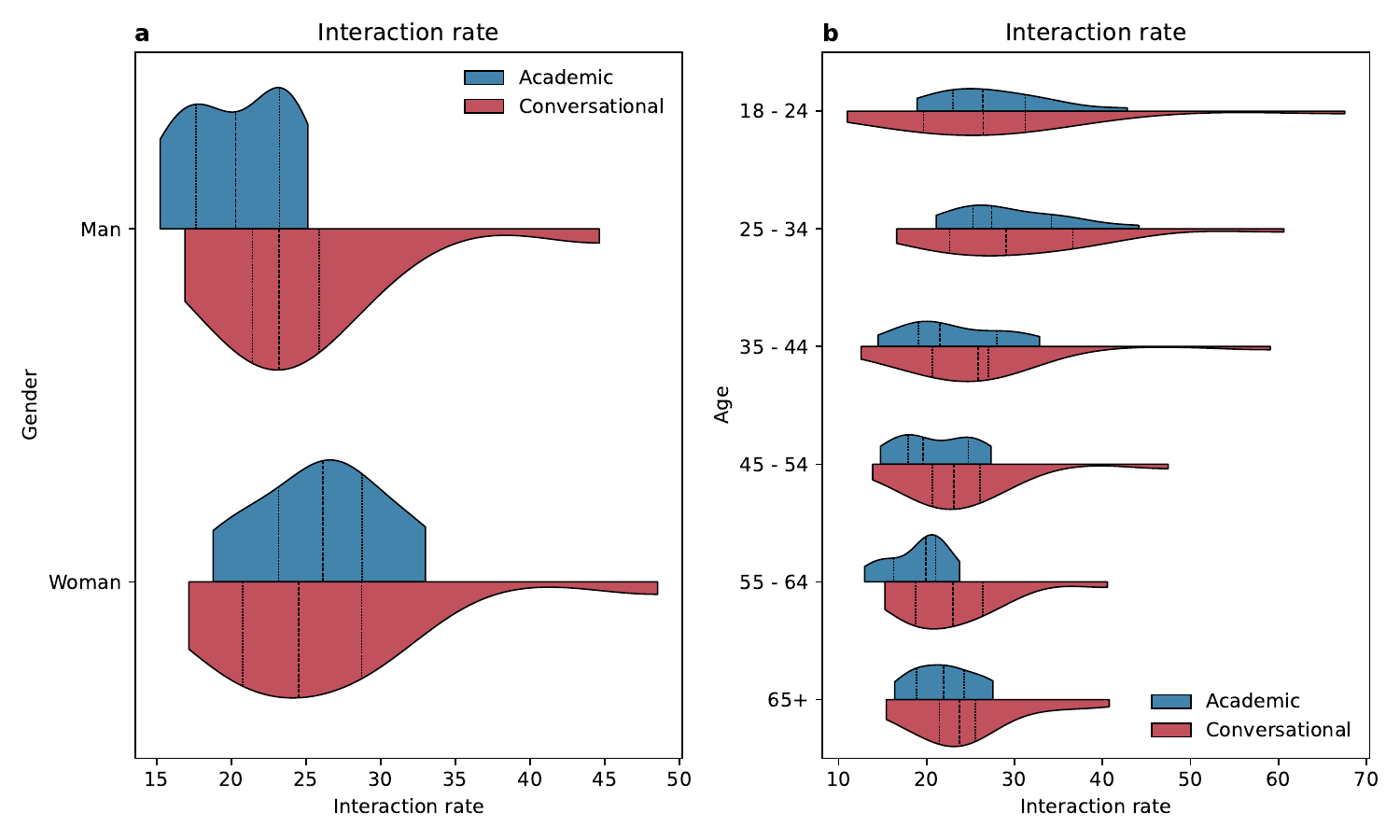}
  \end{center}
\caption{\textbf{Interaction rate by audience profile.} Distribution of interaction rate by (\textbf{a}) gender , (\textbf{b}) age (\textbf{c}) device and (\textbf{d}) the combination of age and gender.}
  \label{fig:adsdistribution_interaction}
\end{figure}

\begin{figure}[!htbp]
  \begin{center}
  \includegraphics[width=0.9\textwidth]{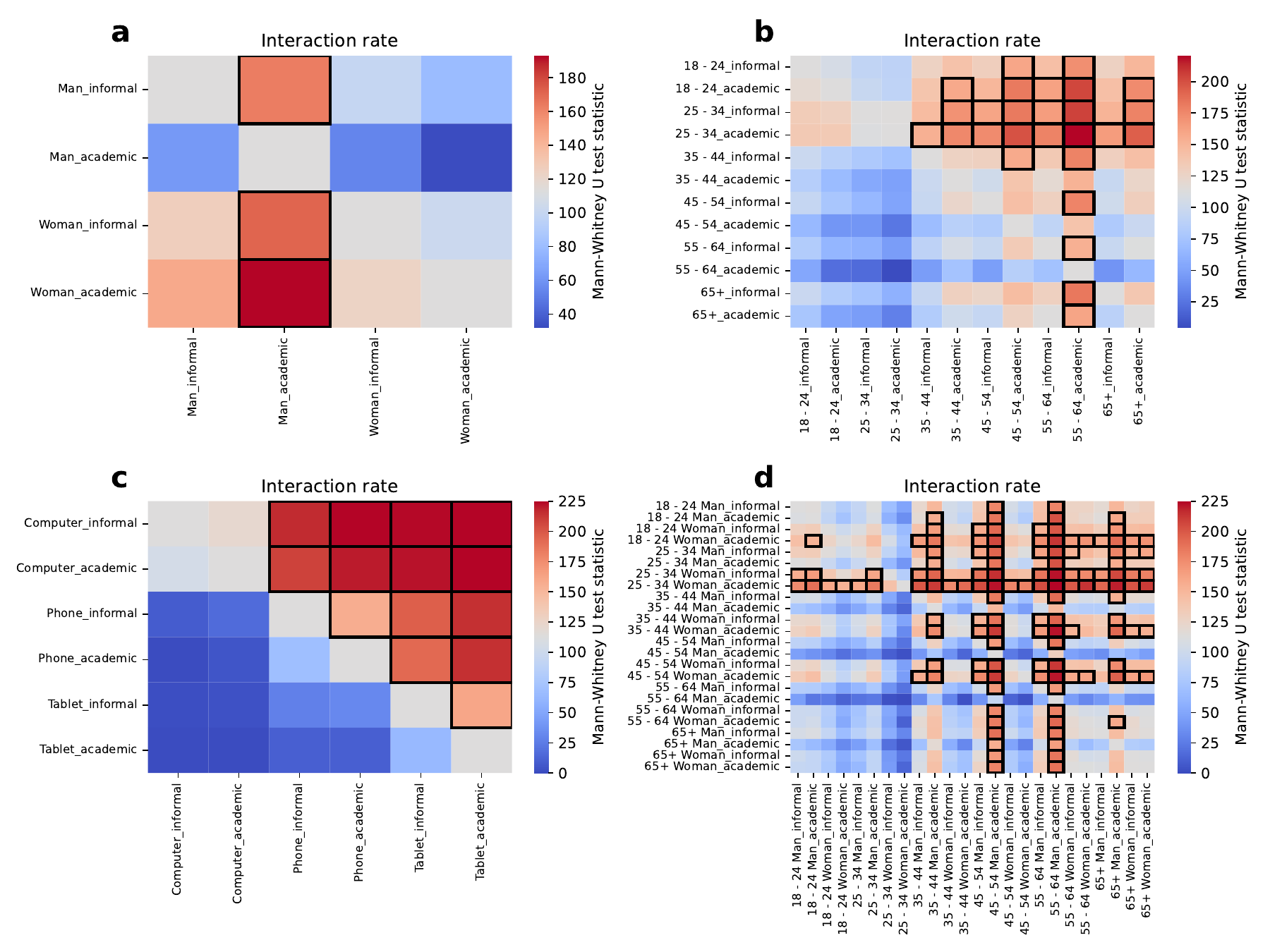}
  \end{center}
\caption{\textbf{Statistical tests between the interaction rate by audience profile.} Results of the statistical tests between the distributions of the interaction rate by (\textbf{a}) gender , (\textbf{b}) age (\textbf{c}) device, and (\textbf{d}) the combination of age and gender. Entries marked in black show statistically significant tests (p-value$<$0.05).}
  \label{fig:adsstatiscal_interaction}
\end{figure}

In Fig. \ref{fig:adsdistribution_interaction}, we show the distribution of the rate of interactions by gender profile, age, device, and combinations of age and gender. The results are similar to the view rates, with the views, women, and young individuals having higher interaction rates compared to men and older persons. The statistical tests reinforce the results (Fig. \ref{fig:adsstatiscal_interaction}), with the interaction rate of men for the academic channel being significantly smaller than the rest.

\begin{figure}[!htbp]

  \begin{center}

  \includegraphics[width=0.9\textwidth]{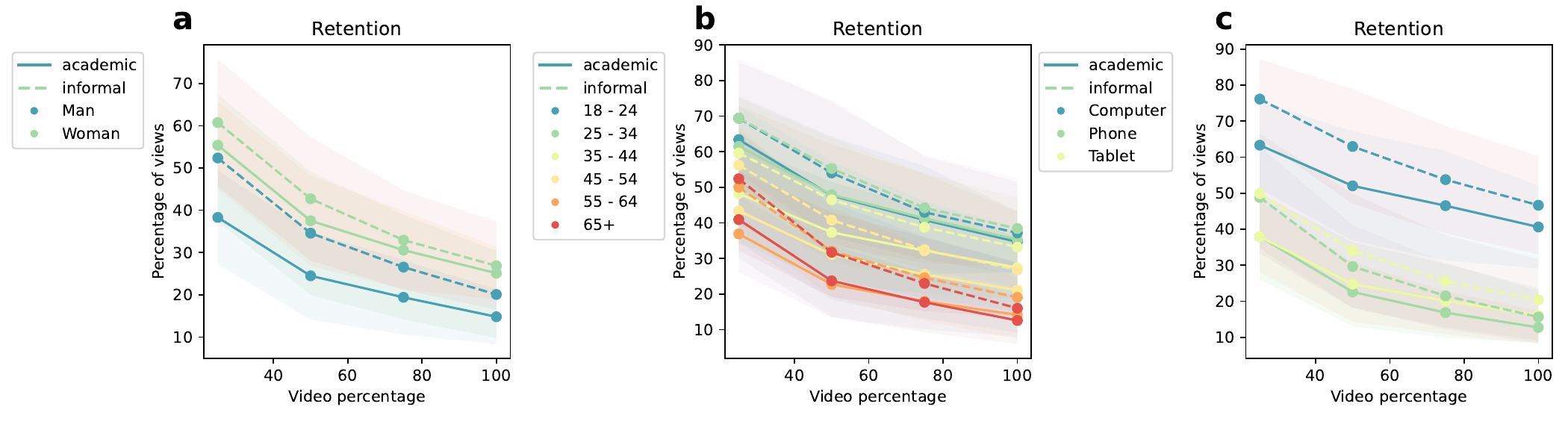}

  \end{center}

\caption{\textbf{Video retention based on audience profile.} Percentage of users viewing 25\%, 50\%, 75\% and 100\% of content by (\textbf{a}) gender , (\textbf{b}) age and (\textbf{c}) device.}

  \label{fig:adsretention}

\end{figure}

In Fig. \ref{fig:adsretention}, we can see what percentage of users have viewed 25\%, 50\%, 75\%, and 100\% of the content, divided by profile and channel. In line with previous results, women show greater retention. For example, only 20\% of men watch the academic videos in their entirety while 30\% of women do. There are also notable differences in age, as 40\% of users in the 25-34 age range view the content entirely, while less than 20\% do so in the age group over 54. Among young individuals, we also see the smallest differences in retention between informal and academic videos. The retention analysis by device reveals that views on computers tend to last longer than on other devices.

\end{document}